%% file: main_final.tex
\newif\ifconfver
\confverfalse      
\confvertrue        

\newif\ifplainver  
\plainvertrue

\newif\ifhide  
\hidetrue

\ifplainver
    \confverfalse   
\fi

\ifconfver
     \documentclass[10pt,twocolumn,twoside]{IEEEtran}
\else
    \ifplainver
        \documentclass[11pt]{article}
        \usepackage{fullpage}
    \else
   \documentclass[12pt,draftcls,onecolumn]{IEEEtran}
    \fi
\fi

\usepackage{etoolbox}%
\usepackage{xpatch}
\usepackage{blindtext}
\usepackage{tocloft}%
\newlength{\articlesectionshift}%
\let\LaTeXStandardSection\section

\let\LaTeXStandardTheParagraph\theparagraph

\makeatletter
\newcounter{titlecounter}

\xpretocmd{\maketitle}{\ifnumgreater{\value{titlecounter}}{1}}{\clearpage}{}{} 
\xpatchcmd{\maketitle}{\let\maketitle\relax\let\@maketitle\relax}{\refstepcounter{titlecounter}\begingroup
  \addtocontents{toc}{\begingroup\addtolength{\cftsecindent}{-\articlesectionshift}}%
  \addcontentsline{toc}{section}{\protect{\numberline{\thetitlecounter}{\@title~ \@author}}}%
  \addtocontents{toc}{\endgroup}
}{%
  \typeout{Patching was successful}
}{%
  \typeout{patching failed}
}%

\def\@IEEEdestroythesectionargument#1{\LaTeXStandardSection{#1}}%

\xapptocmd{\maketitle}{%
\renewcommand{\thesection}{\LaTeXStandardTheSection}%
\renewcommand{\thesubsection}{\LaTeXStandardTheSubSection}%
\renewcommand{\thesubsubsection}{\LaTeXStandardTheSubSubSection}%
\renewcommand{\theparagraph}{\LaTeXStandardTheParagraph}%
}{}{}%

\@addtoreset{section}{titlecounter}

\usepackage{calc,amsfonts,amssymb,amsmath,bm,url,color,theorem,graphicx,cite}
\usepackage{psfrag,float}
\usepackage{algorithm}
\usepackage{algorithmic}
\usepackage{soul}
\usepackage{enumerate}
\usepackage{bbm}
\usepackage{shortcuts_OPT}
\usepackage{multirow}
\usepackage{subcaption}
\usepackage{subfig}
\usepackage[normalem]{ulem}

\usepackage{pifont}%

\ifconfver \else \usepackage{titling} \fi

\usepackage{eqparbox}

\definecolor{orange}{RGB}{255,107,0}

\def\revcolor{\color{black}}  
\def\rev2color{\color{black}} 
\def\revR1color{\color{black}} 

\def\revMovecolor{\color{black}} 

\def\revMoveInsidecolor{\color{black}} 

\def\finalcolor{\color{black}} 

\newtheorem{Lemma}{Lemma}
\newtheorem{Prop}{Proposition}
\newtheorem{Theorem}{Theorem}
\newtheorem{Def}{Definition}

\newtheorem{Asm}{Assumption}
\theorembodyfont{\rmfamily}

\input{sym_marco.tex}

\usepackage{tikz}
\usetikzlibrary{arrows.meta}
\usetikzlibrary{decorations}
\usetikzlibrary{calc}
\usetikzlibrary{shapes.geometric}
\usetikzlibrary{external}

\patchcmd{\maketitle}
 {\def\@makefnmark}
 {\def\@makefnmark{}\def\useless@macro}
 {}{}
\renewcommand\footnotemark{}

\begin{document}


\newcommand{\papertitle}{
Extreme Point Pursuit---Part I: A Framework for Constant Modulus Optimization
}

\newcommand{\paperabstract}{
This study develops a framework for a class of constant modulus (CM) optimization problems, which covers binary constraints, discrete phase constraints, semi-orthogonal matrix constraints, non-negative semi-orthogonal matrix constraints, and several types of binary assignment constraints.
Capitalizing on the basic principles of concave minimization and error bounds, we study a convex-constrained penalized formulation for general CM problems.
The advantage of such formulation is that it allows us to leverage non-convex optimization techniques, such as the simple projected gradient method, to build algorithms.
As the first part of this study, we explore the theory of this framework.
We study conditions under which the formulation provides exact penalization results. 
We also examine computational aspects relating to the use of the projected gradient method for each type of CM constraint.
Our study suggests that the proposed framework has a broad scope of applicability.
}


\ifplainver


    \title{\papertitle}

    \author{
   	Junbin Liu$^\dagger$, Ya Liu$^\dagger$, Wing-Kin Ma$^\dagger$, Mingjie Shao$^\S$ and Anthony Man-Cho So$^\ddagger$\\
     \\
    $^\dagger$ Department of Electronic Engineering, The Chinese University of Hong Kong,\\
     Hong Kong SAR of China \\[.5em]
    $^\ddagger$ Department of Systems Engineering and Engineering Management,\\ The Chinese University of Hong Kong,
    Hong Kong SAR of China\\[.5em] 
    $^\S$School of Information Science and Engineering, \\ Shandong University, Qingdao, China
    \thanks{The work of J. Liu, Y. Liu, W.-K. Ma and M. Shao was supported by the General Research Fund (GRF) of Hong Kong Research Grant Council (RGC) under Project ID CUHK 14208819. The work of A. M.-C. So was supported by the GRF of Hong Kong RGC under Project ID CUHK 14205421.} 
    }  
    \maketitle
    \begin{abstract}
    \paperabstract
    \end{abstract}

\bigskip

\noindent 
{\bf Keywords:} \
        constant modulus optimization,
        non-convex optimization, 
        error bound,
        densest subgraph problem,  PCA,
        graph matching, clustering, ONMF

\else
    \title{\papertitle}

    \ifconfver \else {\linespread{1.1} \rm \fi

    \author{Junbin Liu, Ya Liu, Wing-Kin Ma, Mingjie Shao and Anthony Man-Cho So    
     \thanks{The work of J. Liu, Y. Liu, W.-K. Ma and M. Shao was supported by the General Research Fund (GRF) of Hong Kong Research Grant Council (RGC) under Project ID CUHK 14208819. The work of A. M.-C. So was supported by the GRF of Hong Kong RGC under Project ID CUHK 14205421.} 
    }

    \maketitle

    \ifconfver \else
        \begin{center} \vspace*{-2\baselineskip}
        \end{center}
    \fi

    \begin{abstract}
        \paperabstract
    \end{abstract}


    \begin{IEEEkeywords}\vspace{-0.0cm}
        constant modulus optimization,
        non-convex optimization, 
        error bound,
        densest subgraph problem,  PCA,
        graph matching, clustering, ONMF
    \end{IEEEkeywords}

    \ifconfver \else \IEEEpeerreviewmaketitle} \fi

 \fi

\ifconfver \else
    \ifplainver \newpage \else
\fi \fi



\section{Introduction}

Optimization with constant modulus (CM) constraints appears in a wide variety of applications in signal processing, communications, data science and related fields.
We deal with 
binary and discrete phase constraints in MIMO detection \cite{ma2004semidefinite,shao2020binary,shao2023explanation}, radar code waveform designs \cite{de2008design}, one-bit and constant envelope precoding \cite{shao2019framework}, phase retrieval \cite{waldspurger2015phase}
and phase-only beamforming \cite{tranter2017fast};
semi-orthogonal matrix constraints in various forms of principal component analysis (PCA),
such as sparse PCA \cite{zou2006sparse,journee2010generalized}, robust PCA \cite{kwak2008principal}, fair PCA \cite{tantipongpipat2019multi} and heteroscedastic 
PCA \cite{hong2021heppcat};
binary selection constraints in graph bisection \cite{karisch2000solving} and the densest $k$-subgraph problem \cite{feige2001dense};
permutation matrix constraints in quadratic assignment \cite{burkard2012assignment} and graph matching \cite{zaslavskiy2008path};
size-constrained assignment matrix constraints in paper-to-session assignment \cite{sidiropoulos2015signal} and size-constrained clustering \cite{zhu2010data};
non-negative semi-orthogonal matrix constraints for orthogonal non-negative matrix factorization (ONMF)~\cite{ding2006orthogonal,pompili2014two}.

Optimization with CM constraints is, in general, challenging. 
There has been much enthusiasm with devising practical schemes for tackling, or approximating, CM problems.
The CM constraint structures, perhaps together with the objective function structures, are carefully utilized to build methods in a case-specific fashion.
That led to a vast array of techniques, from combinatorial optimization, convex relaxation, 
to non-convex optimization.
Given the massive volume of literature and the wide variety of methods related to CM optimization,
there are widely different perspectives on how CM problems are treated; e.g., some are concerned with theoretical computational aspects such as NP-hardness and conditions under which the problem is tractable, while some directly use intuitions to build heuristics and apply them to practical applications.
Here we concern ourselves with the non-convex optimization perspective.
In this context we may say that we are well-versed with minimizing a possibly non-convex objective function over a convex set.
Minimization over a non-convex CM set, in comparison, requires much sharper skills in a case-specific way;
examples are
manifold optimization for semi-orthogonal matrix constraints~\cite{chen2020proximal,AMS08,B23} and the generalized power method~\cite{journee2010generalized,LYS23}.
There are also methods that avoid the difficulty of non-convex CM constraints by applying penalization, placing a penalty function on the objective function as ``soft'' constraints~\cite{ding2006orthogonal,pompili2014two,XLY22}.
It is also worthwhile to mention the concave minimization approach \cite{zaslavskiy2008path}, which will be considered in this paper.


Our study has a simple beginning.
We want to extend our prior study \cite{shao2019framework}. 
In that study we considered the cases of binary, discrete-phase and continuous-phase constraints. 
We derived a convex-constrained penalized formulation for the problem at hand, and subsequently we built a projected gradient-based algorithm which was numerically found to work well.
In the current study, we, on the one end, want to look back to uncover the most basic, the most primal, principles that underlie our prior study.
On another end, we want to see how far we can expand the principles to various applications.

With these as our goals, this study sets out to build a convex-constrained penalized framework for a class of CM problems.
Named extreme point pursuit (EXPP), this framework covers a collection of CM constraints, such as the constraints described at the beginning of the Introduction.
At the heart of our study is the aspect of {\em exact penalization}, that is, conditions under which EXPP is an equivalent formulation of its corresponding CM problem.
We will study this aspect under two different principles, namely, the concave minimization and error bound principles.
We will see that EXPP serves as an exact penalization formulation for a wide scope of problems.
Also, like its predecessor, EXPP leads to an equivalent formulation that has a benign structure from the viewpoint of building algorithms; it can be handled by methods as simple as the vanilla projected gradient (or subgradient) method.
In this regard we will examine computational aspects relating to the use of the projected gradient (or subgradient) method for different CM constraints.




We should describe prior related works.
Concave minimization (see, e.g., \cite[Section~32]{rockafellar1970convex}) is a classic notion that has been used in various ways in optimization.
For example, it was described as a way to provide exact penalization in the context of quadratic assignment \cite{bazaraa1982use} back in the 1980's.
The idea is popularly used in the related context of graph matching \cite{zaslavskiy2008path},
{\revR1color and it was also considered in quadratic assignment \cite{xia2010efficient,jiang2016l_p}.}
Our prior study also touched upon concave minimization in an indirect way; see \cite[Theorem 2]{shao2019framework}.
We however have not seen a study that systematically expands the notion of concave minimization to general CM problems.
Moreover, 
error bounds are a notion that has deep roots in constrained optimization for providing exact penalization guarantees \cite{luo1996mathematical,CP22}. 
Nevertheless, we see very few studies on using error bounds to treat CM problems. 
Our prior study may be seen as accidentally stumbling upon the essence of error bounds
in a narrow sense; see \cite[Theorem 1]{shao2019framework}. 
Most recently, error bounds begin to receive attention in 
the context of 
ONMF and related problems \cite{jiang2023exact,chen2022tight}.
Once again, we have not seen a systematic development of error bound techniques for a wide class of  CM problems.


Part of this paper was presented in conferences.
In \cite{shao2022extreme} we presented a premature version of EXPP under the concave minimization principle.
In \cite{liu2024cardinality} we applied EXPP to the 
{\rev2color densest $k$-subgraph}
problem.
The development of error bound techniques, which occupies a significant portion of contributions of our study, is presented for the first time in this paper.

As Part I of this paper, the contents are organized as follows.
Section \ref{sect:prob_stat} provides the problem statement.
Section \ref{sect:concave_min} uses the concave minimization principle to establish EXPP for a class of CM problems.
{\revR1color There, we also examine computational aspects relating to the use of the projected gradient method.}
Section \ref{sect:error_bnds} studies EXPP using the error bound principle and establishes new conditions for exact penalization (which concave minimization cannot cover).
{\revR1color Section \ref{sect:locmin} further covers the aspect of exact penalization with respect to locally optimal solutions.} 
Section \ref{sect:conclusion} concludes our findings.
Part II of this paper will provide numerical results in different applications.
It will also derive new case-specific EXPP results.

\subsection{Notation}

The symbols $\Zbb$, $\Rbb$, $\Rbb_+$, $\Rbb_{++}$, and $\Cbb$ denote the sets of integers, real numbers, non-negative numbers, positive numbers, and complex numbers, respectively.
A vector and a matrix are represented by a boldfaced lowercase letter and a boldfaced capital letter, such as $\bx$ and $\bX$, respectively.
A vector is always assumed to be a column vector in this paper.
The $i$th component of a given vector $\bx$ is denoted by $x_i$.
The $(i,j)$th component of a given matrix $\bX$ is denoted by $x_{ij}$.
The transpose of a given vector $\bx$ is denoted by $\bx^\top$.
The expression $\bx = (x_1,\ldots,x_n)$ is synonymous with $\bx = [~ x_1,\ldots, x_n ~]^\top$.
The trace of a given matrix $\bX \in \Rbb^{n \times n}$ is denoted by $\tr(\bX)$.
Given a matrix $\bX \in \Rbb^{m \times n}$, the vectors $\bx_j \in \Rbb^m$ and $\bar{\bx}_i \in \Rbb^n$ denote the $j$th column and $i$th row of $\bX$, respectively.

The symbols $\bzero$, $\bone$, and $\bI$ denote an all-zero vector,  an all-one vector, and an identity matrix, respectively.
The symbol $\be_i$ denotes a unit vector with $[\be_i]_i= 1$ and $[\be_i]_j = 0$ for all $j \neq i$.
The component-wise absolute value and power-of-$p$ of a given vector $\bx \in \Rbb^n$ are denoted by
$| \bx | = (|x_1|,\ldots,|x_n|)$ and $\bx^p = (x_1^p,\ldots,x_n^p)$, respectively.
The expression $\max\{ \bx, \by \}$ denotes the component-wise  maximum of two vectors $\bx, \by \in \Rbb^n$, i.e., 
$\max\{ \bx, \by \} = (\max\{x_1,y_1\},\ldots,\max\{x_n,y_n\})$.
Given two vectors $\bx, \by \in \Rbb^n$, the inequality $\bx \geq \by$ means that $x_i \geq y_i$ for all $i$.
Given two vectors $\ba, \bb \in \Rbb^n$ with $\ba \leq \bb$,
the symbol $[\bx]_{\ba}^{\bb}:= ( \min\{ \max\{a_1,x_1\},b_1\},\ldots,\min\{ \max\{a_n,x_n\},b_n \} )$ denotes the clipping of a given vector $\bx \in \Rbb^n$ over an interval $[\ba,\bb]$.
The $i$th largest component of a given vector $\bx \in \Rbb^n$ is denoted by $x_{[i]}$.
The function $$s_k(\bx) := x_{[1]} +  {\revcolor \cdots +} x_{[k]}$$ denotes the max-$k$-sum of $\bx \in \Rbb^n$.

Given a scalar $p \geq 1$, 
the $\ell_p$ norm of a vector $\bx \in \Rbb^n$ is denoted by $\| \bx \|_p = ( \sum_{i=1}^n | x_i |^p )^{1/p}$;
and the $\ell_\infty$ norm of $\bx$ is denoted by $\| \bx \|_\infty = \max\{ |x|_1,\ldots, |x|_n \}$.
Similarly, the component-wise $\ell_p$ norm  of a matrix $\bX \in \Rbb^{m \times n}$ (with $p \geq 1$) is denoted by $\| \bX \|_{\ell_p} = ( \sum_{i=1}^m \sum_{j=1}^n | x_{ij} |^p )^{1/p}$.
The $i$th largest singular value of a given matrix $\bX$ is denoted by $\sigma_i(\bX)$.
Given a matrix $\bX \in \Rbb^{m \times n}$, we denote $\bsig(\bX) = ( \sigma_1(\bX),\ldots, \sigma_{r}(\bX))$, where $r= \min\{m,n\}$.
The spectral norm, Frobenius norm, and nuclear norm of a matrix $\bX \in \Rbb^{m \times n}$ are denoted by $\| \bX \|_2 = \sigma_1(\bX)$, $\| \bX \|_{\rm F} = ( \sum_{i=1}^{r} \sigma_i(\bX)^2 )^{1/2} = \| \bX \|_{\ell_2}$, and 
$\| \bX \|_* = \sum_{i=1}^{r} \sigma_i(\bX)$, respectively (where $r= \min\{m,n\}$).

Given a non-empty set $\setX \subseteq \Rbb^n$, the convex hull of $\setX$ is denoted by $\conv(\setX)$.
We denote
\begin{align*}
\Delta^n & = \{ \bx \in \Rbb^n_+ \mid \bone^\top \bx =1 \}, \\
\setB^n & = \{ \bx \in \Rbb^n \mid \| \bx \|_2 \leq 1 \}, \\
\setB^{n,r} & = \{ \bX \in \Rbb^{n \times r} \mid \| \bX \|_2 \leq 1 \},
\end{align*}
as the unit simplex on $\Rbb^n$, the unit $\ell_2$ norm ball on $\Rbb^n$, and the unit spectral norm ball on $\Rbb^{n \times r}$, respectively.
The gradient of a given function $f: \setD \rightarrow \Rbb$, with $\setD \subseteq \Rbb^n$, is denoted by $\nabla f(\bx)$.
The distance from a point $\bx \in \Rbb^n$ to a non-empty set $\setX \subseteq \Rbb^n$ is defined as
$\dist(\bx,\setX) = \inf \{ \| \bx - \bz \|_2 \mid \bz \in \setX \}$.
The projection of a point $\bx \in \Rbb^n$ onto a non-empty set $\setX \subseteq \Rbb^n$ is denoted by $\Pi_\setX(\bx)$;
specifically, $\Pi_\setX(\bx)$ denotes any point in $\setX$ such that $\| \bx - \Pi_\setX(\bx) \|_2 = \dist(\bx,\setX)$.

The above notation applies to matrices and complex-valued vectors, whenever applicable.
Also, we denote $\jj = \sqrt{-1}$.
The real component, imaginary component, and angle of a complex scalar $x$ are denoted by $\Re(x)$, $\Im(x)$, and $\angle x$, respectively.

\section{Problem Statement}
\label{sect:prob_stat}

\subsection{The Problem}

In this study, we consider a class of constant modulus (CM) problems.
To describe it, 
a set $\setV \subset \Rbb^n$ is said to be a CM set with modulus $\sqrt{C}$ if $\| \bx \|_2^2 = C$ for any $\bx \in \setV$.
We focus on problems that take the form 
\beq \label{eq:cm}
\min_{\bx \in \setV} f(\bx),
\eeq 
where $f: \setD \rightarrow \Rbb$ is a function with domain $\setD \subseteq \Rbb^n$; $\setV \subseteq \setD$ 
is a non-empty closed CM set with modulus $\sqrt{C}$ and with $\conv(\setV) \subseteq \setD$.
We will place particular emphasis on the following CM sets.
\begin{enumerate}[1.]
	
	\item {\em Binary vector set:}  $\{ -1, 1 \}^n$.
	
	\item {\em $m$-ary phase shift keying (MPSK) set:} 
    $$\Theta_m:= \{ x \in \Cbb \mid x= e^{\jj \frac{2\pi l}{m} + \jj \frac{\pi}{m} }, ~ l \in \{0,1,\ldots,m-1\} \},$$
    where $m \geq 3$ is an integer.
	
	\item {\em Unit sphere:} $$\setS^n := \{ \bx \in \Rbb^n \mid \| \bx \|_2 = 1 \}.$$
    An example is the complex unit-modulus set $\{ x \in \Cbb \mid |x|= 1 \}$.
	
	\item {\em Semi-orthogonal matrix set:}
	$$\setS^{n,r}:= \{ \bX \in \Rbb^{n \times r} \mid \bX^\top \bX = \bI \},$$
	where $n \geq r$.

	\item {\em Unit vector set:} $$\setU^n := \{ \be_1,\ldots,\be_n \} \subseteq \Rbb^n.$$
	
	\item {\em Selection vector set:}
	$$\setU_\kappa^n  :=  
	\{ \bx \in \{0,1\}^n \mid \bone^\top \bx = \kappa \},$$
	where $\kappa \in \{1,\ldots,n\}$.

	\item {\em Partial permutation matrix set:}
	\begin{align*}
	\setU^{n,r} & := 
	\{ \bX \in \Rbb^{n \times r} \mid \bx_j \in \setU^n ~ \forall j, ~\bx_j^\top \bx_{j'} = 0 ~ \forall j \neq j' \},	 
	\end{align*}
	where $n \geq r$. 
	Note that 
    $\setU^{n,n}$
    is the full permutation matrix set.
	The set $\setU^{n,r}$ can be characterized as 
	\[
	\setU^{n,r} = \{ \bX \in \{ 0, 1 \}^{n \times r} \mid  \bX^\top \bone = \bone, \bX \bone \leq \bone \},
	\]
	and, for $n = r$, it can be further written as 
	\[
	\setU^{n,n} = \{ \bX \in \{ 0, 1 \}^{n \times n} \mid  \bX^\top \bone = \bone, \bX \bone = \bone \}.
	\]
	
	\item {\em Size-constrained assignment matrix set:}
	\begin{align}
	\setU^{n,r}_{\bka}
	& := \{ \bX \in \Rbb^{n \times r} \mid 
	\bx_j \in \setU_{\kappa_j}^n ~ \forall j, ~ \bx_j^\top \bx_{j'} = 0 ~ \forall j \neq j' \}, 
	\nonumber 
	\end{align} 
	where $n \geq r$, $\bka \in \{1,\ldots,n\}^r$, $\sum_{j=1}^r \ka_j \leq n$.
	The set $\setU^{n,r}_{\bka}$ can be characterized as 
	\[
	\setU^{n,r}_{\bka} = \{ \bX \in \{ 0, 1 \}^{n \times r} \mid  \bX^\top \bone = \bka, \bX \bone \leq \bone \},
	\]
	and, for $\sum_{j=1}^r \ka_j = n$, it can be further written as 
	\[ 	\label{eq:set_Qk}
	\setU^{n,r}_\bka = \{ \bX \in \{ 0, 1 \}^{n \times n} \mid  \bX^\top \bone = \bka, \bX \bone = \bone \}.
	\]


	\item {\em Non-negative semi-orthogonal matrix set:} 
	$$\setS^{n,r}_+ := \setS^{n,r} \cap \Rbb_+^{n \times r},$$ where $n \geq r$.

    \item {\em Cartesian product of CM sets:}
    $\setV = \setV_1 \times \cdots \times \setV_r$, where each $\setV_i \subseteq \Rbb^{n_i}$ is a CM set with modulus $\sqrt{C}_i$.
    Examples are the MPSK vector set $\Theta_m^n$ and the complex component-wise unit-modulus set $\{ \bx \in \Cbb^n \mid |\bx| = \bone \}$.
	
\end{enumerate}
Note that the unit sphere $\setS^n$ is a special case of the semi-orthogonal matrix set $\setS^{n,r}$;
the unit vector set $\setU^n$ is a special case of the selection vector set $\setU^n_\ka$;
the selection vector set $\setU^n_\ka$ and the partial permutation matrix set $\setU^{n,r}$
are special cases of the size-constrained assignment matrix set $\setU_\bka^{n,r}$.
The reader is referred to the beginning of the Introduction and the references therein for the applications.



\subsection{The Approach to Be Pursued}

There are many different approaches to deal with specific CM problems. 
For instance, when the constraint set $\setV$ is a smooth manifold in an Euclidean space (such as the unit sphere $\setS^n$ or the semi-orthogonal matrix set $\setS^{n,r}$), the CM problem~\eqref{eq:cm} can be tackled by manifold optimization techniques~\cite{AMS08,B23}. On the other hand, if the constraint set $\setV$ can be expressed as a subgroup of the orthogonal group (such as the MPSK set $\Theta_m$, the unit sphere $\setS^n$, the semi-orthogonal matrix set $\setS^{n,r}$, or the full permutation matrix set $\setU^{n,n}$), then it can also be tackled by the so-called generalized power method; see, e.g., \cite{LYS23} and cf.~\cite{journee2010generalized}. However, the efficiency of these approaches depends crucially on that of computing the projection onto the often {\it non-convex} constraint set $\setV$. Moreover, these approaches cannot tackle the entire class of CM problems that we are interested in.
This study aims to provide a {\em unified} treatment of a large class of CM problems 
using penalization techniques. Specifically, we
want to convert the general CM problem  \eqref{eq:cm} into a convex-constrained minimization problem
\beq \label{eq:cm_pen}
\min_{\bx \in \setX} f(\bx) + \lambda \, h(\bx),
\eeq 
where $\setX \subseteq \setD$ is a constraint set that is convex and closed,
$h:\Rbb^n \rightarrow \Rbb$ is a penalty function that promotes the decision variable $\bx$ to lie in the CM set $\setV$,
and $\lambda > 0$ is a given scalar.
To ensure that such a conversion is advantageous from both theoretical and numerical viewpoints, our goal is to find an $\setX$ and an $h$ such that (i) problem \eqref{eq:cm_pen} is an equivalent formulation of the CM problem  \eqref{eq:cm}, in the sense that there is a correspondence between the (locally) optimal solutions to the two problems; (ii) problem \eqref{eq:cm_pen} can be tackled by standard numerical methods that enjoy convergence guarantees and are practically efficient.

\subsection{Remarks}

We want to discuss a basic aspect relating to classic constrained optimization. 
Readers who want to immediately see our methods can jump to the next section.
Advanced readers from optimization will notice that the formulation \eqref{eq:cm_pen} is the same as that by the classic penalty methods in constrained optimization; see, e.g., \cite[Chapter 17]{nocedal2000} or \cite[Chapter 9]{CP22}. This is true, but there is also a difference.
In the context of constrained optimization, it is common to assume that $\setX = \setD$.
Or, the chosen $\setX$ has little relationship with the structure of the original constraint set $\setV$.
Moreover, the penalty functions arising from classic constrained optimization could be unfriendly from the viewpoint of building algorithms.
For example, for the binary case $\setV = \{ -1, 1\}^n$, the application of the quadratic penalty method in constrained optimization \cite[Chapter 17]{nocedal2000} leads to the following penalty function
\[
h(\bx) = \| \bone - | \bx |^2 \|_2^2.
\]
As the reader will see, we take inspiration from basic optimization principles to derive both $\setX$ and $h$ for exact penalization results.
For example, for the binary case we will see that
\[
h(\bx) = - \| \bx \|_2^2, \quad \setX= [-1,1]^n
\]
gives exact penalization results. Our approach can be viewed as a concrete instantiation of the theory of exact penalization in \cite[Chapter 9]{CP22} and yields, for the first time, exact penalization results that are specifically tailored for CM problems.

\section{Exact Penalization by Concavification}
\label{sect:concave_min}

In this section, we study the notion of concave minimization for converting a CM problem to a convex-constrained minimization problem.
In the first subsection, we will first review the basic idea by using  MIMO detection as an example.
Then, in the second subsection, we will expand the idea as a principle for general CM problems. 
The third subsection will introduce an optimization scheme under the principle to be established, and the fourth subsection will assess the applicability and computational issues of the resulting scheme for various CM sets.
The fifth subsection will summarize the development and discuss further aspects.

\subsection{Example: MIMO Detection}

As an example, consider the following problem 
\beq \label{eq:mimo_det}
\min_{\bx \in \{-1,1\}^n } f(\bx)= 
 \| \by - \bH \bx \|_2^2
\eeq 
for some given $\by \in \Rbb^m$ and $\bH \in \Rbb^{m \times n}$.
This problem is known as the MIMO detection problem in the context of communications and signal processing.
The reader is referred to the literature (e.g., the references in \cite{shao2020binary}) for the background.
Here we focus on a reformulation of problem \eqref{eq:mimo_det} as a convex-constrained minimization problem.
Given a scalar $\lambda > 0$, consider the following penalized formulation
\beq \label{eq:mimo_det_nsp}
\min_{\bx \in [-1,1 ]^n } F_\lambda(\bx):= f(\bx) - 
\lambda 
\| \bx \|_2^2.
\eeq 
This formulation was proposed in \cite{shao2019framework}.
The idea is to use a negative square penalty  to force each variable $x_i$ to have its modulus $|x_i|$ as large as possible, 
which, under the constraint $x_i \in [-1,1]$, should lead $x_i$ to be close to either $-1$ or $1$.
It can be shown that, for a sufficiently large $\lambda$, 
the penalized formulation \eqref{eq:mimo_det_nsp} is an equivalent formulation of the MIMO detection problem \eqref{eq:mimo_det} \cite[Theorem 2]{shao2019framework}.
Consider the following lemma.
\begin{Lemma} \label{lem:cvx_min}
	Let $\setD \subseteq \Rbb^n$.
	Let $\setA \subseteq \setD$ be a non-empty set.
	Let $\Phi: \setD \rightarrow \Rbb$ be a function that is strictly concave at least on $\conv(\setA)$, with $\conv(\setA) \subseteq \setD$.
	Suppose that 
	\beq \label{eq:gen_prob1}
	\min_{\bx \in \setA} \Phi(\bx)
	\eeq 
	has an optimal solution.
	Then the following problem
	\beq \label{eq:gen_prob1_cvx}
	\min_{\bx \in \conv(\setA)} \Phi(\bx)
	\eeq 
	is an equivalent formulation of problem \eqref{eq:gen_prob1} in the sense that
	(a) problem \eqref{eq:gen_prob1_cvx} has an optimal solution, given by any one of the optimal solutions to \eqref{eq:gen_prob1};
	and (b) any optimal solution to \eqref{eq:gen_prob1_cvx} must be an optimal solution to \eqref{eq:gen_prob1}.
\end{Lemma}

Lemma \ref{lem:cvx_min} is an elementary result in optimization \cite[Section~32]{rockafellar1970convex}.
Since its proof is easy to understand and provides insight, we concisely review the proof.

\medskip
{\em Proof of Lemma \ref{lem:cvx_min}:} \
Let $\bx$ be any point in $\conv(\setA)$.
By definition, we can represent $\bx$ by
$\bx = \sum_{i=1}^k \theta_i \ba_i$ for some
$\ba_1,\ldots,\ba_k \in \setA$, $\btheta \in \Rbb^k_+$, $\sum_{i=1}^k \theta_i = 1$, and $k$.
Let $\bx^\star$ denote any optimal solution to \eqref{eq:gen_prob1}.
We have
\begin{subequations} \label{eq:pf:cvx_min_eq1}
	\begin{align}
	\Phi(\bx) & \geq \textstyle \sum_{i=1}^k \theta_i \Phi(\ba_i) 
	\label{eq:pf:cvx_min_eq1a} \\
	& \geq \textstyle \sum_{i=1}^k \theta_i \Phi(\bx^\star)  
	\label{eq:pf:cvx_min_eq1b} \\
	& = \Phi(\bx^\star),
	\end{align}
\end{subequations}
where \eqref{eq:pf:cvx_min_eq1a} is due to Jensen's inequality and the concavity of $\Phi$,
and \eqref{eq:pf:cvx_min_eq1b} is due to $\Phi(\ba) \geq \Phi(\bx^\star)$ for any $\ba \in \setA$.
Eq.~\eqref{eq:pf:cvx_min_eq1} implies that the minimum of $\Phi$ over $\conv(\setA)$ can be attained, and it is attained at $\bx^\star$.
In other words, problem \eqref{eq:gen_prob1_cvx} has an optimal solution, and any optimal solution to \eqref{eq:gen_prob1} is an optimal solution to \eqref{eq:gen_prob1_cvx}.
Moreover, \eqref{eq:pf:cvx_min_eq1a} attains equality if and only if $\btheta = \be_j$ for some $j$; this is because $\Phi$ is strictly concave.
This implies that the minimum of $\Phi$ over $\conv(\setA)$ is attained at $\bx$ only if $\bx \in \setA$,
which 
implies that any optimal solution to \eqref{eq:gen_prob1_cvx} is also an optimal solution to \eqref{eq:gen_prob1}.
The proof is complete.
\hfill $\blacksquare$
\medskip

It is known that $\conv(\{-1,1\}^n) =  [-1,1]^n$. 
By expanding $$F_\lambda(\bx) = \bx^\top ( \bH^\top \bH - \lambda \bI ) \bx - 2 \by^\top \bH^\top \bx + \| \by \|_2^2,$$
we see that $F_\lambda$ is strictly concave when $\bH^\top \bH - \lambda 
\bI $ is negative definite.
Suppose that 
$\lambda > \sigma_1(\bH)^2$
such that $\bH^\top \bH - \lambda 
\bI $ is negative definite.
Then Lemma \ref{lem:cvx_min} asserts that problem \eqref{eq:mimo_det_nsp} is an equivalent formulation of
\[
\min_{\bx \in \{-1,1\}^n} F_\lambda(\bx) = \min_{\bx \in \{-1,1\}^n} f(\bx) - 
\lambda n
\]
which is the MIMO detection problem \eqref{eq:mimo_det}.
Hence, given any $\lambda >  \sigma_1(\bH)^2$,
problem \eqref{eq:mimo_det_nsp} is an equivalent formulation of the MIMO detection problem \eqref{eq:mimo_det}.
To provide the reader with the intuition, we graphically illustrate the optimization landscape in Fig.~\ref{fig:illus_F_lambda}(a).
We see that as $F_\lambda$ becomes concave, it exhibits a down-slope landscape.
Such landscape causes the optimal $x$ to go to either $-1$ or $1$.


\subsection{Extreme Point Pursuit for General CM Problems}

We want to expand on the idea in the previous subsection to general CM problems.
Given a scalar $\lambda > 0$,
consider the following penalized formulation of the CM problem \eqref{eq:cm}: 
\beq \label{eq:expp}
\min_{\bx \in 	\conv(\setV)} F_\lambda(\bx):= f(\bx) - 
\lambda
\| \bx \|_2^2.
\eeq 
From the exact penalization result in Lemma \ref{lem:cvx_min},
we know that problem \eqref{eq:expp} is an equivalent formulation of the CM problem \eqref{eq:cm} if $F_\lambda$ is strictly concave.
We have learned in the previous example that if $f$ is quadratic, we can always concavify the penalized objective function $F_\lambda$.
Now the question is how far we can go beyond quadratic functions. 

To answer this question, we consider the notion of weak convexity.
Let us provide the context.
Let $\setD \subseteq \Rbb^n$.
Given a scalar $\rho > 0$ and a convex set $\setX \subseteq \setD$, a function $\phi: \setD \rightarrow \Rbb$ is said to be $\rho$-weakly convex on $\setX$ if $\phi(\bx) + \rho \| \bx \|^2_2/2$ is convex on $\setX$.
A popular subclass of weakly convex functions is the class of functions that are differentiable and have the Lipschitz continuous gradient property.
Given a scalar $L > 0$ and a set $\setX \subseteq \setD$, a function $\phi: \setD \rightarrow \Rbb$ is said to have $L$-Lipschitz continuous gradient on $\setX$ if $\phi$ is differentiable and 
$\| \nabla \phi(\bx) - \nabla \phi(\bx') \|_2 \leq L \| \bx - \bx' \|_2$ for all $\bx,\bx' \in \setX$.
We have the following result.
\begin{Lemma}{\bf(\cite{drusvyatskiy2018proximal})} \label{lem:cvx_push0}
	If $\phi: \setD \rightarrow \Rbb$ has $L$-Lipschitz continuous gradient on a convex set $\setX \subseteq \setD$, then $\phi$ is $L$-weakly convex on $\setX$.
    In particular, given any $\mu > L$, the function $\phi(\bx) + \mu \| \bx \|_2^2/2$ is strictly convex on $\setX$.
\end{Lemma}
In our study, we want $f$ to be weakly concave; i.e., $-f$ is weakly convex. By definition, if $\phi$ has $L$-Lipschitz continuous gradient on $\setX$, then $-\phi$ also has $L$-Lipschitz continuous gradient on $\setX$.
This leads to the following important variant of Lemma~\ref{lem:cvx_push0}.
\begin{Lemma} \label{lem:cvx_push}
	If $\phi: \setD \rightarrow \Rbb$ has $L$-Lipschitz continuous gradient on a convex set $\setX \subseteq \setD$, then $-\phi$ is $L$-weakly convex on $\setX$.
    In particular, given any $\mu > L$, the function $\phi(\bx) - \mu \| \bx \|_2^2/2$ is strictly concave on $\setX$.
\end{Lemma}

We are now ready to provide a sufficient condition under which the penalized formulation \eqref{eq:expp} gives exact penalization results for general CM problems. 
Assume the following.
\begin{Asm} \label{asm:Lip}
	The objective function $f$ of the CM problem \eqref{eq:cm} has $L$-Lipschitz continuous gradient on $\conv(\setV)$.
\end{Asm}
Assumption~\ref{asm:Lip} is considered applicable in many applications.
 This is because any twice {\revcolor continuously} differentiable function $f$ 
has Lipschitz continuous gradient on a compact set, 
and $\conv(\setV)$ is always compact.
The following theorem describes the exact penalization result.
\begin{Theorem} \label{thm:cvx_min}
	Consider the CM problem in \eqref{eq:cm}.
	Suppose that Assumption~\ref{asm:Lip} holds.
	Given any scalar $\lambda > L/2$, problem \eqref{eq:expp} is an equivalent formulation of the CM problem \eqref{eq:cm} in 
	the sense that the optimal solution sets of problems \eqref{eq:expp} and \eqref{eq:cm} are equal.
\end{Theorem}
As a remark, it is even possible to show that the {\em locally} optimal solution sets of problems \eqref{eq:cm} and \eqref{eq:expp} are equal for many CM sets of interest; 
{\revR1color this will be described later in Section~\ref{sect:locmin}.}

The proof of Theorem~\ref{thm:cvx_min} is as follows. By Lemma~\ref{lem:cvx_push}, the function $F_\lambda$ in \eqref{eq:expp} is strictly concave when $\lambda > L/2$.
Applying Lemma \ref{lem:cvx_min} gives the desired result.

\begin{figure*}[htb]
     \centering
     \begin{subfigure}[]{0.32\textwidth}
         \centering
         \includegraphics[width=\textwidth]{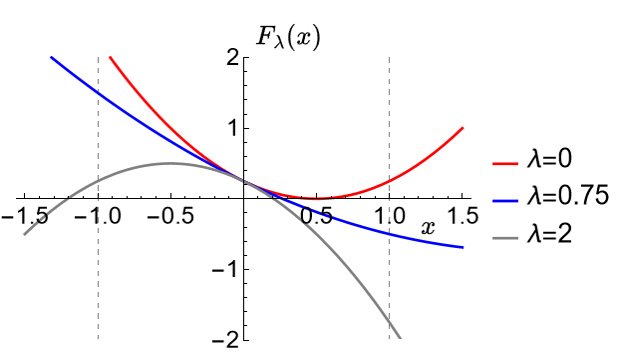}
         \caption{\scalebox{0.8}{$f(x) = (x-0.5)^2$.}}
         \label{fig:illustration_order2}
     \end{subfigure}
     \hfill
     \begin{subfigure}[]{0.32\textwidth}
         \centering
         \includegraphics[width=\textwidth]{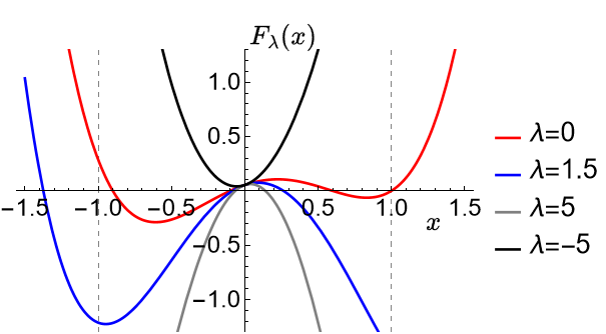}
         \caption{\scalebox{0.8}{$f(x)=(x + 0.9)(x + 0.1)(x - 0.6)(x - 1)$.}}
         \label{fig:illustration_order4}
     \end{subfigure}
     \hfill
     \begin{subfigure}[]{0.32\textwidth}
         \centering
         \includegraphics[width=\textwidth]{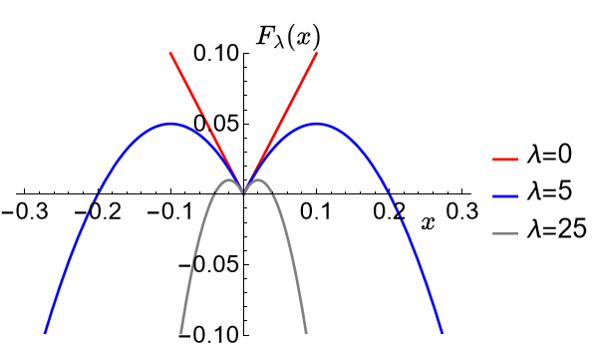}
         \caption{\scalebox{0.8}{$f(x)=|x|$.}}
         \label{fig:illustration_abs}
     \end{subfigure}
\caption{Illustration of the concavification effects with $F_\lambda(\bx) = f(\bx) - \lambda \| \bx \|_2^2$. (a) A convex quadratic instance of $f$; (b) a non-convex differentiable instance of $f$; (c) a non-differentiable instance of $f$.}
\label{fig:illus_F_lambda}
\end{figure*}

Theorem~\ref{thm:cvx_min} asserts that, given a sufficiently large scalar $\lambda > 0$, the penalized formulation \eqref{eq:expp} is an equivalent formulation of a general CM problem.
In what follows, we will call the penalized formulation \eqref{eq:expp} the  {\em extreme point pursuit (EXPP) problem} of the CM problem \eqref{eq:cm}.
This is because every point in a CM set $\setV$ can be shown to be an extreme point of $\setV$, and the penalty $-\| \bx \|_2^2$ has the flavor of pulling the minimizer of \eqref{eq:expp} to approach an extreme point.
Fig.~\ref{fig:illus_F_lambda}(b) graphically illustrates such effects.




\subsection{Projected Gradient Methods and Homotopy Optimization} \label{subsec:pgm}

Our next question is how we deal with the EXPP problem \eqref{eq:expp} in terms of building algorithms to find its solution. 
The EXPP problem \eqref{eq:expp} is an optimization problem with a convex constraint and a non-convex differentiable objective function;
the objective function $F_\lambda$ also possesses the Lipschitz continuous gradient property.
Let us suppose that it is easy to compute $\Pi_{\conv(\setV)}(\cdot)$, the projection onto $\conv(\setV)$;
this is a key issue and we will come back to this in the next subsection.
A naturally suitable class of optimization methods for the EXPP problem would be the projected gradient method. Specifically, consider
\beq \label{eq:pgd}
\bx^{l+1} = \Pi_{\conv(\setV)}( \bx^l - \eta_l \nabla F_\lambda(\bx^l) ),
\quad l=0,1,\ldots,
\eeq 
where $\eta_l > 0$ is the step size, 
the vector $\bx^l$, $l \geq 1$, denotes the $l$th iterate of the algorithm,
and $\bx^0$ denotes a given starting point for the algorithm.
It is well-known in the context of first-order optimization that, if $1/\eta_l$ is chosen to be no less than the Lipschitz constant of the gradient of $F_\lambda$, then the projected gradient algorithm in \eqref{eq:pgd} has several convergence properties with respect to a critical point or a near-critical point of the EXPP problem \eqref{eq:expp};
see, e.g., \cite[Chapter 10]{beck2017first}, \cite{ABS13} (particularly Theorem 5.3) and the references therein.

We would like to consider another aspect, which is based on our previous empirical experience in MIMO detection \cite{shao2020binary} and was also indicated in a related study \cite{zaslavskiy2008path}.
The exact penalization result in Theorem~\ref{thm:cvx_min} tells us to use a large scalar $\lambda$ for the EXPP problem.
If we choose a large $\lambda$ and then run an algorithm (such as the projected gradient algorithm) for the EXPP problem \eqref{eq:expp}, we found that the results are usually poor.
It seems that the algorithm can easily get trapped in poor local minima when $\lambda$ is large.
In practice, we often employ a homotopy optimization strategy \cite{shao2020binary,zaslavskiy2008path,dunlavy2005homotopy,wu1996effective,hazan2016graduated}.
We start with a small $\lambda$
such that the corresponding EXPP problem \eqref{eq:expp} is ``easy'' to solve.
Specifically, if the objective function $f$ is convex, we can start with $\lambda = 0$ and
the corresponding EXPP problem \eqref{eq:expp} is a convex optimization problem.
If $f$ is non-convex, we can still make the problem easy.
By Lemma~\ref{lem:cvx_push0}, we know that, under Assumption~\ref{asm:Lip}, the function $F_\lambda$ is convex if $\lambda < -L/2 $;
that is, we can convexify $F_\lambda$.
We illustrate the convexification effects in Fig.~\ref{fig:illus_F_lambda}(b); see the curve for $\lambda= -5$.
By starting with $\lambda < -L/2 $, we start with a convex problem.
What we do next is to gradually increase $\lambda$.
The intuition is that the optimization landscape should change gradually as we change $\lambda$ gradually.
Hence, by starting with an easy EXPP problem and then tackling the EXPP problems with gradually increasing $\lambda$'s, we might be able to trace the optimal solution path of the EXPP problem with respect to $\lambda$, thereby gradually approaching the optimal solution to the CM problem.
So far there is no guarantee that homotopy optimization can find the optimal solution,
but empirical results look promising.
We describe the homotopy optimization method as a pseudo-code in Algorithm~\ref{alg:hot}.

\begin{algorithm}[!t]
	\caption{A homotopy optimization method.} \label{alg:hot}
	\begin{algorithmic}[1]
		
		\STATE \textbf{given:} a sequence $\{ \lambda_k \}$ and a starting point $\bx^0$
		
		\STATE $k \leftarrow 0$
		
		\REPEAT 
		
		\STATE run an algorithm for problem \eqref{eq:expp} with $\lambda = \lambda_k$ (e.g., the projected gradient algorithm in \eqref{eq:pgd}), with $\bx^k$ as the starting point.
		Set $\bx^{k+1}$ as the solution obtained by the algorithm. 
		
		\STATE $k \leftarrow k+1$
		
		\UNTIL{a stopping rule is met}
		
		\STATE \textbf{output:} $\bx^k$

	\end{algorithmic}
\end{algorithm}


\subsection{Are Projections Onto Convex Hulls Easy?}
\label{sect:cvx_hull_proj}

When we discuss the optimization aspects in the last subsection,
we make a simplifying assumption that it is easy to compute $\Pi_{\conv(\setV)}(\cdot)$.
Now we return to this aspect by examining our interested CM sets described in Section~\ref{sect:prob_stat}.

\begin{figure}[htb!]
    \centering
\includegraphics[scale=0.5]{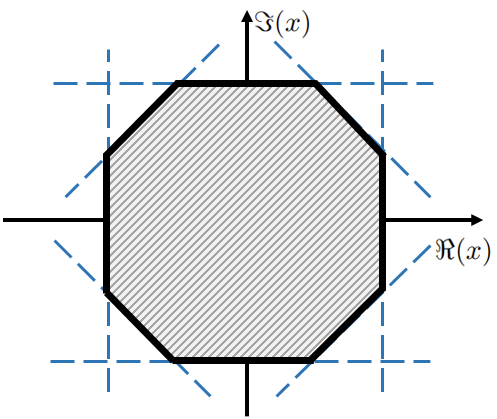}
    \caption{The convex hull of the MPSK set. $m= 8$.}
    \label{fig:mpsk-convexhull}
\end{figure}

\begin{enumerate}[1.]
	\item {\em Binary vector set:} 
	We have $\conv(\{-1,1\}^n)= [-1,1]^n$ and $\Pi_{[-1,1]^n}(\bz) = [ \bz ]_{-\bone}^{\bone}$.
	
	\item {\em MPSK set:} 
    Given an integer $m \geq 3$, define 
    \ifconfver
        $$ 
        \begin{aligned}
        \setP_m  =  \Big\{ x \in \Cbb & \, \Big| \, \Re \left(e^{\jj \frac{2\pi l}{m}} x \right) \leq \cos\left( \tfrac{\pi}{m} \right), l \in \{0,\ldots,m-1\} \Big\}.
        \end{aligned}
	$$
    \else
     $$ 
        \setP_m =  \big\{ x \in \Cbb \, \big| \, \Re \left(e^{\jj \frac{2\pi l}{m}} x \right) \leq \cos\left( \tfrac{\pi}{m} \right), ~ l \in \{0,\ldots,m-1\} \big\}. 
    $$
    \fi 
    This set equals $\conv(\Theta_m)$.
    An illustration of $\setP_m$ is shown in 
    Fig.~\ref{fig:mpsk-convexhull}.
    The projection onto $\setP_m$ is 
    \[
        \Pi_{\setP_m}(z) = e^{\jj \frac{2\pi k}{m}} \left( [\Re(y)]_0^{\cos(\pi/m)} + \jj [\Im(y) ]_{-\sin(\pi/m)}^{\sin(\pi/m)}  \right),
    \] 
    where $k = \left\lfloor (\angle z + \pi/m)/(2\pi/m) \right\rfloor$, $y = z e^{-\jj \frac{2\pi k}{m}}$.
	
	\item {\em Unit sphere:} 
	The convex hull of the unit sphere is the unit $\ell_2$ norm ball:	
	 $$\conv(\setS^n)= \{ \bx \in \Rbb^n \mid \| \bx \|_2 \leq 1 \}= \setB^n.$$
    The projection onto $\setB^n$ is
    $\Pi_{\setB^n}(\bz) = \bz$ if $\| \bz \|_2 \leq 1$, and $\Pi_{\setB^n}(\bz) = \bz/\| \bz \|_2$ if $\| \bz \|_2 > 1$.
	
	\item {\em Semi-orthogonal matrix set:} 
	To describe the convex hull, we first note that the semi-orthogonal matrix set can be characterized as
	\[
	\setS^{n,r} = \{ \bX \in \Rbb^{n \times r} \mid \sigma_i(\bX) = 1  ~ \forall i \}.
	\]
	The convex hull of $\setS^{n,r}$ is the unit spectral norm ball:
	\beq
	\nonumber \\
	\conv(\setS^{n,r})= \{ \bX \in \Rbb^{n \times r} \mid \sigma_i(\bX) \leq 1 ~ \forall i \}
	= \setB^{n,r}.
	\eeq 
	To describe the projection onto $\setB^{n,r}$, let $\bZ \in \Rbb^{n \times r}$ be a given matrix.
	Let $\bZ = \bU \bSig \bV^\top$ be its singular value decomposition (SVD), where $\bU \in \Rbb^{n \times r}$ is semi-orthogonal, $\bSig= \Diag(\bm \sigma(\bZ))$,
	and $\bV \in \Rbb^{r \times r}$ is orthogonal.
    We have 
    $$\Pi_{\setB^{n,r}}(\bZ)= \bU \Diag([ \bm \sigma(\bZ)]_{\bzero}^{\bone}) \bV^\top.$$
	The computational cost with the projection is mainly with the SVD, which requires $\mathcal{O}(n r^2)$ operations.

	\item {\em Unit vector set:}
	The convex hull of the unit vector set is the unit simplex:
	\beq  
	\nonumber 
	\conv(\setU^n) = \{ \bx \in \Rbb_+^n \mid \bone^\top \bx = 1 \} = \Delta^n.
	\eeq 
	The projection onto $\Delta^n$ does not have a closed form, but it can be computed by an algorithm with a complexity of $\bigO( n \log(n) )$ \cite{condat2016fast}.
		
	\item {\em Selection vector set:} 
	It can be shown that 
	\beq \label{eq:conv_sel}
	\conv(\setU_\kappa^n) = \{ \bx \in [0,1]^n \mid \bone^\top \bx = \kappa \}.
	\eeq 
	The projection onto $\conv(\setU_\kappa^n)$ can be computed by a bisection search algorithm \cite[Algorithm 1]{konar2021exploring} with a complexity of $\bigO( n \log(n/\varepsilon))$, where $\varepsilon > 0$ describes the solution precision.

	\item {\em Partial permutation matrix set:}
	The Birkhoff--von Neumann theorem asserts that the convex hull of the full permutation matrix set is the set of doubly stochastic matrices:
	\[
	\conv(\setU^{n,n})= \{ \bX \in [0,1]^{n\times n} \mid \bX^\top \bone = \bone, \bX \bone = \bone \}.
	\]
    For the general case $n \geq r$, it can be shown that
	\[
	\conv(\setU^{n,r})= \{ \bX \in [0,1]^{n \times r} \mid \bX^\top \bone = \bone, \bX \bone \leq \bone \}.
	\]
    {\revR1color The projection onto $\conv(\setU^{n,r})$ does not have a closed-form or easy-to-compute solution, to the best of our knowledge.
    A numerical solver is required to solve the projection problem.
    One convenient option is to use a general-purpose convex optimization software, such as CVX \cite{cvx}.
    Another option is to find a specialized solver, one that exploits the problem structure for efficient computation.
    One such solver is the dual gradient method in  \cite{jiang2016l_p}; it was developed for the case of $n=r$ and can be extended to the case of $n \geq r$.
    This method has a complexity per iteration of  $\bigO(nr)$, but it may take many iterations to reach an accurate result.
    In summary, while it is possible to numerically compute the projection onto $\conv(\setU^{n,r})$, it may be practically expensive to do so particularly when $n$ and $r$ are large. 
    }

	\item {\em Size-constrained assignment matrix set:}
	It can be shown that
	\begin{align*}
	\conv(\setU^{n,r}_{\bm \kappa})
	&  = \{ \bX \in [ 0,1 ]^{n \times r} \mid \bX^\top \bone = \bm \kappa, \bX \bone \leq \bone \}.
	\end{align*}
    {\revR1color 
    The projection onto $\conv(\setU^{n,r}_{\bka})$ 
    faces the same computational issues discussed in the last point. 
    It is a more general problem, and thus no easier, than its counterpart for $\conv(\setU^{n,r})$.
    One can find a numerical solver to compute the projection onto $\conv(\setU^{n,r}_{\bka})$, but that may be expensive particularly for large $n$ and $r$.}

	\item {\em Non-negative semi-orthogonal matrix set:}
	To the best of our knowledge, the expression of the convex hull of $\setS^{n,r}_+ = \setS^{n,r} \cap \Rbb_+^{n \times r}$ is not known.
	We are unable to derive it, either.
	It is very tempting to think that $\conv(\setS^{n,r}_+) = \conv(\setS^{n,r}) \cap \Rbb_+^{n \times r} = \setB^{n,r} \cap \Rbb_+^{n \times r}$.
	Unfortunately this is not true: 
	the point $\bzero$ lies in $\setB^{n,r} \cap \Rbb_+^{n \times r}$,
	but any point in $\conv(\setS_+^{n,r})$ can be shown to be nonzero.
	Given our lack of knowledge of 
    $\conv(\setS^{n,r}_+)$,
    we do not know what is the associated projection.

    \item {\em Cartesian product of CM sets:}
    As a basic fact,
    \[
    \conv(\setV_1 \times \cdots \times \setV_r) =
    \conv(\setV_1) \times \cdots \times \conv(\setV_r);
    \]
    see \cite[Problem~1.37]{bertsekas2003convex}.
    Given a vector $\bz = (\bz_1,\ldots,\bz_r)$, $\bz_i \in \Rbb^{n_i}$, the projection of $\bz$ onto  $\conv(\setV_1) \times \cdots \times \conv(\setV_r)$ is $( \Pi_{\conv(\setV_1)}(\bz_1),\ldots,\Pi_{\conv(\setV_r)}(\bz_r) )$.
	
\end{enumerate}

We provide the proof of some of the main convex hull and projection results in \ifconfver {\revcolor the supplemental material.}
\else{\revcolor Appendix \ref{appendix:cvx_hull&proj}. }\fi
While some results are known, e.g., \cite[Section 3.4]{journee2010generalized} for $\setS^{n,r}$ and \cite[Theorem 9.8.3]{brualdi2006combinatorial} for $\setU^{n,r}$ (see also \cite{LYS23} for related results), there are also results that 
we could not find accurate proofs from the literature.

\subsection{Summary and Further Remarks}

Let us pause and give a mini summary.
We studied a convex-constrained penalized formulation \eqref{eq:expp}, called EXPP, for a general class of CM problems \eqref{eq:cm}.
The EXPP formulation has a benign structure, and it can be handled by the projected gradient method as a universal algorithmic scheme.
The computational efficiency of implementing the projected gradient method is however case-specific. 
It crucially depends on whether the projection onto the convex hull of the CM set is easy to compute.
Our study showed that 
\begin{enumerate}[(a)]
    \item the cases of binary vectors, MPSK vectors, unit sphere vectors, semi-orthogonal matrices, unit vectors and selection vectors have efficient-to-compute projections;
    \item the cases of partial/full permutation matrices and size-constrained assignment matrices have computable projections, but the costs 
    {\rev2color can be}
    high for large $n$ and $r$;
    \item the case of non-negative semi-orthogonal matrices is an open problem, 
and we do not know how to apply EXPP.
\end{enumerate}
In conclusion, 
EXPP
is considered ``easy'' to use for CM constraints
under category (a).
The other CM constraints will be further studied in 
our subsequent development.

\section{Exact Penalization by Error Bounds}
\label{sect:error_bnds}

The core principle of the EXPP formulation in the last section is to turn $F_\lambda(\bx) = f(\bx) - \lambda \| \bx \|_2^2$ to a concave function.
This concave minimization principle can cover a broad class of differentiable objective functions $f$'s, but it may not cover non-smooth $f$.
As an example, consider $f(\bx) = \| \bx \|_1$.
It can be verified that $F_\lambda(\bx)$ can never be concave, regardless of how large $\lambda$ is; see Fig.~\ref{fig:illus_F_lambda}(c) for an illustration.

In this section, we study the EXPP formulation under the error bound principle \cite{luo1996mathematical},
seeing if the EXPP formulation can actually provide exact penalization results for a class of possibly non-smooth objective functions.
The first subsection will first review elementary error bound results. 
In the second subsection, we will consider an example, namely, one-bit MIMO precoding, to provide insights into how the error bound principle interacts with EXPP.
The third and fourth subsections will link the error bound principle with EXPP for the general case and specific cases, respectively.
The fifth and sixth subsections will provide technical proofs.
The seventh subsection will give further discussion.

\subsection{Elementary Results}
\label{sect:eb_elementary}

First we introduce the notion of Lipschitz continuity.
Let $\setD \subseteq \Rbb^n$.
Given a scalar $K > 0$ and a set $\setX \subseteq \setD$,
a function $\phi: \setD \rightarrow \Rbb$ is said to be $K$-Lipschitz continuous on $\setX$ if 
\[
| \phi(\bx) -  \phi(\bx') | \leq K \| \bx - \bx' \|_2,
\quad \forall \bx,\bx' \in \setX.
\]
For example, $\phi(\bx) = \| \bx \|_1$ is $\sqrt{n}$-Lipschitz continuous on $\Rbb^n$.
A quadratic function is {\em not} Lipschitz continuous on $\Rbb^n$, but it is Lipschitz continuous on a compact set.


Second we describe basic constrained optimization results that utilize Lipschitz continuity to achieve exact penalization.
The following result is considered most elementary.
\begin{Lemma}
	{\bf (Proposition 2.4.3 in \cite{clarke1990optimization})}
	\label{lem:ep1}
	Let $\setD \subseteq \Rbb^n$.
	Let $\setA \subseteq \setD$ be a non-empty closed set.
	Let $\phi: \setD \rightarrow \Rbb$ be a function that is $K$-Lipschitz continuous on some set $\setC \subseteq \setD$, with $\setA \subseteq \setC$.
	Suppose that 
	\beq \label{eq:gen_prob2}
	\min_{\bx \in \setA} \phi(\bx)
	\eeq 
	has an optimal solution.
	Given any scalar $\lambda > K$, the following problem
	\beq \label{eq:gen_prob2_ep}
	\min_{\bx \in \setC} \phi(\bx) + \lambda \, \dist(\bx,\setA)
	\eeq 
	is an equivalent formulation of problem \eqref{eq:gen_prob2} in the sense that
	(a) problem \eqref{eq:gen_prob2_ep} has an optimal solution, given by any one of the optimal solutions to \eqref{eq:gen_prob2};
	and
	(b) any optimal solution to \eqref{eq:gen_prob2_ep} must be an optimal solution to \eqref{eq:gen_prob2}. 
\end{Lemma}
%

The proof of Lemma \ref{lem:ep1} is easy to understand.
Given its importance,
we review the proof to give readers insight.

\medskip
{\em Proof of Lemma \ref{lem:ep1}:}
Let $\bx$ be any point in $\setC$.
As an elementary property of $\dist(\cdot,\setA)$ with non-empty closed $\setA$, there exists a point $\ba \in \setA$ such that $\| \bx - \ba \|_2 = \dist(\bx,\setA)$ \cite{clarke1990optimization}.
Let $\Phi(\bx) = \phi(\bx) + \lambda \, \dist(\bx,\setA)$.
Let $\bx^\star \in \setA$ denote any optimal solution to \eqref{eq:gen_prob2}.
We have 
\begin{subequations} \label{eq:pf:ep1_eq1}
	\begin{align}
	\Phi(\bx) & \geq \textstyle \phi(\boldsymbol{a}) + (\lambda - K) \| \bx - \ba \|_2 
	\label{eq:pf:ep1_eq1a} \\
	& \geq \phi(\bx^\star)  = \Phi(\bx^\star),
	\label{eq:pf:ep1_eq1b} 
	\end{align}
\end{subequations}
where \eqref{eq:pf:ep1_eq1a} is due to the $K$-Lipschitz continuity of $\phi$ relative to $\setC$,
and \eqref{eq:pf:ep1_eq1b} is due to $\lambda > K$ and $\phi(\ba) \geq \phi(\bx^\star)$ for any $\ba \in \setA$.
Eq.~\eqref{eq:pf:ep1_eq1} implies that $\Phi$ attains a minimum over $\setC$ at $\bx^\star$.
Hence, problem \eqref{eq:gen_prob2_ep} has an optimal solution, and any optimal solution to \eqref{eq:gen_prob2} is an optimal solution to \eqref{eq:gen_prob2_ep}.
Moreover, \eqref{eq:pf:ep1_eq1b} attains its equality only if $\bx = \ba \in \setA$; note that $\lambda > K$.
This implies that any optimal solution to  \eqref{eq:gen_prob2_ep} must also be an optimal solution to \eqref{eq:gen_prob2}.
The proof is complete.
\hfill $\blacksquare$
\medskip

Lemma \ref{lem:ep1} shows that a penalized formulation with the distance function $\dist(\cdot,\setA)$ as the penalty function can lead to an equivalent formulation of the original constrained problem.
Lemma \ref{lem:ep1} provides the foundation for the  error bound principle for exact penalization.
To describe it,
consider the following definition.
\begin{Def}
	Given a set $\setC \subseteq \Rbb^n$ and a set $\setA \subseteq \setC$, a function $\psi: \Rbb^n \rightarrow \Rbb$ is said to be an error bound function of $\setA$ relative to $\setC$ if  
	 \begin{subequations} \label{eq:err_bnd_def}
	 	\begin{align} \label{eq:err_bnd_def_a}
	 	\dist(\bx,\setA) \leq \psi(\bx), & \quad \forall \bx \in \setC, \\
	 	\psi(\bx) = 0, & \quad \forall \bx \in \setA.
	 	\label{eq:err_bnd_def_b}
	 	\end{align}
	 \end{subequations}
 	Alternatively, given an inequality in the form of \eqref{eq:err_bnd_def_a}, we call \eqref{eq:err_bnd_def_a} an error bound relative to $\setC$ if \eqref{eq:err_bnd_def_a} attains equality when $\bx \in \setA$.
\end{Def}

An error bound function can be seen as a majorant of $\dist(\cdot,\setA)$ at $\bzero$ and on $\setC$.
As the successor of Lemma \ref{lem:ep1}, the following result is important and frequently-used. 
\begin{Lemma}
	{\bf 
	(Theorem 2.1.2 and Remark 2.1.3 in \cite{luo1996mathematical}, or 
    {\rev2color Proposition 9.1.1}
    in \cite{CP22})	
	}
	\label{lem:ep2}
	Consider the same setting in Lemma \ref{lem:ep1}.
	Let $\psi$ be an error bound function of $\setA$ relative to $\setC$.	
	Given any scalar $\lambda > K$, the following problem
	\beq \label{eq:gen_prob2_ep2}
	\min_{\bx \in \setC} \phi(\bx) + \lambda \, \psi(\bx)
	\eeq 
	is an equivalent formulation of problem \eqref{eq:gen_prob2} in the same sense as that in Lemma \ref{lem:ep1}. 
\end{Lemma}
We remark that it is also possible to establish a relationship between the {\em locally} optimal solutions to problems~\eqref{eq:gen_prob2} and~\eqref{eq:gen_prob2_ep2}; see, 
{\rev2color e.g.,~\cite[Proposition 9.1.2]{CP22}} {\revR1color and also Section~\ref{sect:locmin}.}

Lemma \ref{lem:ep2} can be straightforwardly derived by following the proof of Lemma \ref{lem:ep1}, and 
we shall not provide the proof in this paper;
see the aforementioned references for the proof.
The advantage of Lemma \ref{lem:ep2} over its predecessor Lemma \ref{lem:ep1} is that it provides us with an opportunity to derive friendly penalty functions.
In the next subsection we will give an example to illustrate why this is so.
Sometimes we may have to make a compromise by accepting inexact penalization.
We derive the following 
result,
which will be useful later.
\begin{Lemma}
	\label{lem:ep3}
	Consider the same setting in Lemmas \ref{lem:ep1}--\ref{lem:ep2}.
	Let $\bx^\star$ denote any optimal solution to \eqref{eq:gen_prob2}.
	Suppose that, given any scalar $\lambda > 0$, the following problem 
	\beq \label{eq:gen_prob2_ep3}
	\min_{\bx \in \setC} \phi(\bx) + \lambda \, \psi(\bx)^2
	\eeq 
	has an optimal solution.
	Let  $\hat{\bx}$ denote any optimal solution to \eqref{eq:gen_prob2_ep3},
	and let $\bx' =  \Pi_{\setA}(\hat{\bx})$ be any projection of $\hat{\bx}$ onto $\setA$. 
	Problem \eqref{eq:gen_prob2_ep3} 
	is an inexact formulation of problem \eqref{eq:gen_prob2} in the sense of the following assertions:
	\begin{enumerate}[(a)]
	\item $\dist(\hat{\bx},\setA) \leq K/\lambda$,
	\item $\phi(\hat{\bx}) \leq \phi(\bx^\star) \leq \phi(\hat{\bx}) + K^2/\lambda$, and 
	\item $\phi(\bx^\star) \leq \phi(\bx') \leq \phi(\bx^\star)  + K^2/\lambda$.	
	\end{enumerate}
\end{Lemma}
The proof of Lemma \ref{lem:ep3} is shown in Appendix \ref{app:proof_lem:ep3}.
Lemma \ref{lem:ep3} suggests that 
\eqref{eq:gen_prob2_ep3} approximates \eqref{eq:gen_prob2} better
as $\lambda$ increases.
In particular, by letting $\varepsilon = \bigO(1/\lambda)$, a solution $\hat{\bx}$ to \eqref{eq:gen_prob2_ep3} is $\varepsilon$-feasible to problem \eqref{eq:gen_prob2} and achieves an $\varepsilon$-close objective value relative to the optimal value of problem \eqref{eq:gen_prob2};
a rounded solution $\bx'$ from a solution to \eqref{eq:gen_prob2_ep3} is $\varepsilon$-optimal to \eqref{eq:gen_prob2}.

\subsection{Example: One-Bit MIMO Precoding}
\label{sect:ep2:1b_mimo_precode}

To illustrate the potential of the exact penalization results in the last subsection,
consider the following example
\beq \label{eq:1b_mimo_precode}
\min_{\bx \in \{-1,1\}^n } f(\bx):= \max_{i=1,\ldots,m} \ba_i^\top \bx + b_i,
\eeq 
where $\ba_1,\ldots,\ba_m \in \Rbb^n$ and $b_1,\ldots,b_m \in \Rbb$ are given.
This problem arises in one-bit MIMO precoding \cite{shao2019framework}.
The objective function $f$ is non-smooth.
It can be verified that $f$ is Lipschitz continuous on $\Rbb^n$, with Lipschitz constant $K= \max_{i=1,\ldots,m} \| \ba_i \|_2$.

We consider applying the exact penalization results in Lemmas \ref{lem:ep1}--\ref{lem:ep2} to problem \eqref{eq:1b_mimo_precode}.
We begin with the more elementary one, Lemma \ref{lem:ep1}.
It is easy to verify that $\dist(\bx,\{-1,1\}^n) = \| \bone - | \bx | \|_2$,
and by Lemma \ref{lem:ep1} we have the following equivalent formulation of problem \eqref{eq:1b_mimo_precode}:
\[
\min_{\bx \in \setX} f(\bx) + \lambda \| \bone - | \bx | \|_2
\]
where $\setX\subseteq \Rbb^n$ is any set that covers $\{ -1, 1 \}^n$.
However, the penalty function with the above formulation does not look too friendly.
We thus turn to error bounds.
Recall that in EXPP we choose 
$\setX = [-1,1]^n$.
For any $\bx \in [-1,1]^n$, we have 
\begin{subequations} \label{eq:exa_binary_errbnd}
	\begin{align}
	\| \bone - | \bx | \|_2 & \leq \| \bone - | \bx | \|_1 
	\label{eq:exa_binary_errbnd_a} \\
	& = \textstyle  \sum_{i=1}^n (1 - | x_i |) 
	\label{eq:exa_binary_errbnd_b} \\
	& \leq n - \textstyle \sum_{i=1}^n | x_i |^2,
	\label{eq:exa_binary_errbnd_c}
	\end{align}
\end{subequations}
where 
\eqref{eq:exa_binary_errbnd_c} is due to $z \geq z^2$ for all $z \in [0,1]$;
also, Eq.~\eqref{eq:exa_binary_errbnd} attains equality when $\bx \in \{ -1, 1 \}^n$.
This means that $n - \| \bx \|_2^2$ is an error bound function of $\{-1,1\}^n$ relative to $[-1,1]^n$.
Applying this error bound to Lemma \ref{lem:ep2} gives the following problem
\beq \label{eq:1b_mimo_precode_expp1}
\min_{\bx \in [-1,1]^n} f(\bx) - \lambda \| \bx \|_2^2 
\eeq 
as an equivalent formulation of problem \eqref{eq:1b_mimo_precode}.
Interestingly, the above formulation is the same as the EXPP problem \eqref{eq:expp}.
This reveals that EXPP can also deal with a non-smooth objective function $f$.
We remark that the error bound-based proof  \eqref{eq:exa_binary_errbnd} is very simple, and it appears to be new.

We give two further remarks.
First, the function $n- \| \bx \|_2^2$ is not the only error bound. 
We can see from \eqref{eq:exa_binary_errbnd} that $n- \| \bx \|_1$ is also an error bound function of $\{-1,1\}^n$ relative to $[-1,1]^n$.
This means that the following formulation
\beq \label{eq:1b_mimo_precode_expp2}
\min_{\bx \in [-1,1]^n} f(\bx) - \lambda \| \bx \|_1 
\eeq 
is also an equivalent formulation of problem \eqref{eq:1b_mimo_precode}.
Second, \eqref{eq:1b_mimo_precode_expp1} and \eqref{eq:1b_mimo_precode_expp2} were already shown to be equivalent formulations of the one-bit MIMO precoding problem \eqref{eq:1b_mimo_precode} in \cite[Theorem 1]{shao2019framework} and  \cite[Theorem 3]{wu2023efficient}, respectively, though they were not shown using error bounds.

\subsection{Error Bounds for General CM Problems and EXPP}

The above example raises a question:
Does the notion of error bounds for exact penalization apply to EXPP in general?
To pose the question more accurately, recall the EXPP problem \eqref{eq:expp} of the general CM problem \eqref{eq:cm}:
\beq  
\label{eq:expp_restate}
\min_{\bx \in 	\conv(\setV)} F_\lambda(\bx)= f(\bx) - 
\lambda
\| \bx \|_2^2.
\eeq
Also, consider the following assumption.
\begin{Asm} \label{asm:Lip2}
	The objective function $f$ of the CM problem \eqref{eq:cm} is $K$-Lipschitz continuous on $\conv(\setV)$.
\end{Asm}
Assumption~\ref{asm:Lip2} is considered reasonable. 
Suppose that the domain $\setD$ of $f$ is open.
It is known that, if $f$ can be written as $f = f_1 + f_2$,
where $f_1$ is {\revcolor continuously} differentiable (not necessarily convex) and $f_2$ is convex (not necessarily differentiable) on $\setD$, then $f$ is Lipschitz continuous on any compact set ($\conv(\setV)$ in our study).
In particular, if $f$ satisfies the Lipschitz gradient assumption in Assumption~\ref{asm:Lip}, then it satisfies Assumption~\ref{asm:Lip2} trivially;
if $f$ is weakly convex, then it satisfies Assumption~\ref{asm:Lip2}.


Our question is whether $C- \| \bx \|_2^2$, or its scaled counterpart, serves as an error bound function of $\setV$ relative to $\conv(\setV)$.
If the answer is yes, then, by the virtue of Lemma \ref{lem:ep2}, the EXPP problem is an equivalent formulation of {\em any} CM problem.
The following theorem provides the answer.

\begin{Theorem} \label{thm:err_bnd_gen}
Let $\setV$ be {\revR1color an arbitrary} CM set {\revR1color with modulus $\sqrt{C}$}.
	\begin{enumerate}[(a)]
		\item There does not exist a bounded constant $\nu > 0$ such that, 
        {\revR1color given any CM set $\setV$, the following inequality}
		\[
		\dist(\bx,\setV) \leq \nu (C- \| \bx \|_2^2), 
		\quad 
		\forall \bx \in \conv(\setV)
		\]
        {\revR1color can be satisfied.}
		In particular, consider the following counter-example.
		Let $\setV = \{ e^{-\jj \varphi}, e^{\jj \varphi} \}$,
		where $\varphi \in (0,\pi/2]$.
		It holds that
		\beq
		\dist(\bx,\setV) \geq \frac{1}{2\sin(\varphi)}(1- | x |^2).
		\eeq

		\item 
        {\revR1color Given any CM set $\setV$, we have an error bound}
		\[
		\dist(\bx,\setV) \leq \sqrt{C - \| \bx \|_2^2},
		\quad \forall \bx \in \conv(\setV).
		\]
		
	\end{enumerate}
\end{Theorem}

The proof of Theorem~\ref{thm:err_bnd_gen}(a) is given in Appendix \ref{app:thm:err_bnd_gen:a}.
The proof of Theorem~\ref{thm:err_bnd_gen}(b) is shown as follows.

\medskip
{\em Proof of Theorem~\ref{thm:err_bnd_gen}(b):} \
Let $\bx$ be any point in $\conv(\setV)$,
and represent it by $\bx = \sum_{i=1}^k \theta_i \bv_i$ for some $\bv_1,\ldots,\bv_k \in \setV$, $\btheta \in \Rbb^k_+$, $\sum_{i=1}^k \theta_i = 1$, and $k$.
We have
\[
\dist(\bx,\setV)^2 \leq \| \bv_i - \bx \|_2^2 
= C - 2 \bv_i^\top \bx + \| \bx \|_2^2,
\]
for all $i$. It follows that
\ifconfver
{\revcolor \begin{align*}
 	\dist(\bx,\setV)^2 & \leq  \textstyle \sum_{i=1}^k \theta_i \| \bv_i - \bx \|_2^2 \\
 	& =  \textstyle C - 2 ( \sum_{i=1}^k \theta_i \bv_i )^\top \bx + \| \bx \|_2^2 \\
 	& = C - \| \bx \|_2^2.
 \end{align*}}
\else
{\revcolor
\[
	\dist(\bx,\setV)^2  \leq  \textstyle \sum_{i=1}^k \theta_i \| \bv_i - \bx \|_2^2  =  \textstyle C - 2 ( \sum_{i=1}^k \theta_i \bv_i )^\top \bx + \| \bx \|_2^2  = C - \| \bx \|_2^2.
\]
}
\fi
The proof is complete.
\hfill $\blacksquare$
\medskip

Theorem~\ref{thm:err_bnd_gen}(a) suggests that, under the error bound principle in Lemma \ref{lem:ep2}, it is impossible to show that the EXPP problem is always an equivalent formulation of a CM problem.
Theorem~\ref{thm:err_bnd_gen}(b) gives rise to the following conclusions.
\begin{Theorem} \label{thm:err_bnd_gen_result}
	Consider the CM problem in \eqref{eq:cm}.
	Suppose that Assumption~\ref{asm:Lip2} holds.
	\begin{enumerate}[(a)]
		\item Given any scalar $\lambda > 0$,
		the EXPP problem in \eqref{eq:expp} or in \eqref{eq:expp_restate} is an inexact formulation of the CM problem \eqref{eq:cm} in the sense of the assertions in Lemma \ref{lem:ep3}.
	
		\item 	Given any scalar $\lambda > K$, 
		the following problem
		\beq
		\label{eq:expp_cor1}
		\min_{\bx \in \conv(\setV)} f(\bx) + 
		\lambda
		\sqrt{C- \| \bx \|_2^2}
		\eeq 
		is an equivalent formulation of the CM problem \eqref{eq:cm} in 
		the sense that the optimal solution sets of problems \eqref{eq:expp_cor1} and \eqref{eq:cm} are equal.
	\end{enumerate}
\end{Theorem}
Theorem \ref{thm:err_bnd_gen_result} is the direct consequence of applying Theorem~\ref{thm:err_bnd_gen} to Lemmas \ref{lem:ep2}--\ref{lem:ep3}.
Theorem \ref{thm:err_bnd_gen_result}(a) indicates that, given a general CM set, EXPP may still provide good approximation results when $\lambda$ is large relative to the Lipschitz constant $K$ of the objective function $f$.
Theorem \ref{thm:err_bnd_gen_result}(b) suggests a fix for achieving exact penalization.
However, the penalty function with the fix in \eqref{eq:expp_cor1} appears to be unfriendly.

\subsection{Error Bounds for Specific CM Sets}

While it is impossible to show that the EXPP problem in \eqref{eq:expp_restate} provides exact penalization results for any CM set, we can consider specific CM sets.
Encouragingly, we found that most of our interested cases do lead to the desired results.
Here is the highlight of what we will show.

\begin{enumerate}[1.]
	\item {\em Binary vector set:} 
	This was shown in Section \ref{sect:ep2:1b_mimo_precode}:
	For any $\bx \in [-1,1]^n$, we have the error bounds
	\begin{align*}
	\dist(\bx,\{-1,1\}^n) \leq n - \| \bx \|_1 \leq n - \| \bx \|_2^2.
	\end{align*}
	
	\item {\em MPSK set:} 
	For any $x \in \conv(\Theta_m)$, we have the error bound
	\begin{align} \label{eq:err_mpsk}
        \dist(x,{\rev2color\Theta_m})
        \leq 
	\nu
	(1 - | x |^2),
	\end{align}
	where $\nu = 2$ for $m=3$, and $\nu = 1/\sin(\pi/m)$ for $m \geq 4$.
	
	\item {\em Unit sphere:} 
	For any $\bx \in \setB^n$, we have the error bounds
	\begin{align} \label{eq:err_unit_sphere}
	\dist(\bx,\setS^n) \leq 1 - \| \bx \|_2 \leq 1 - \| \bx \|_2^2.
	\end{align}
	
	\item {\em Semi-orthogonal matrix set:} 
	For any $\bX \in \setB^{n,r}$, we have the error bounds
	\begin{align} \label{eq:err_S}
	\dist(\bX,\setS^{n,r}) \leq r - \| \bX \|_* \leq r - \| \bX \|_{\rm F}^2.
	\end{align}
	
	\item {\em Unit vector set:}
	For any $\bx \in \Delta^n$ we have the error bounds
	\begin{align} \label{eq:err_U}
	\dist(\bx,\setU^n) \leq 2( 1 - x_{[1]}) \leq 2( 1 - \| \bx \|_2^2).
	\end{align}
	Recall that $x_{[i]}$ denotes the $i$th largest component of $\bx$.
	
	\item {\em Selection vector set:} 
	For any $\bx \in [0,1]^n$ with $\bone^\top \bx = \kappa$, we have the error bounds 
	\begin{align} \label{eq:err_Uk}
	\dist(\bx,\setU^n_\kappa) \leq 2( \kappa - s_\kappa(\bx)) \leq 2( \kappa - \| \bx \|_2^2).
	\end{align}
	Recall that $s_\kappa(\bx) = x_{[1]} + \cdots + x_{[\kappa]}$.
	
	\item {\em Partial permutation matrix set:}
	For any $\bX \in [0,1]^{n \times r}$ with $\bX^\top \bone = \bone$ and $\bX \bone \leq \bone$, we have the error bounds  
	\begin{subequations} \label{eq:error_Unk}
		\begin{align}
		\dist(\bX,\setU^{n,r}) & \leq 3 \sqrt{r} 
		\left[ \sum_{j=1}^r (1 - s_1(\bx_j) ) \right]  \\
		& \leq 3 \sqrt{r} ( r - \| \bX \|_{\rm F}^2 ).
		\end{align}
	\end{subequations}
		
	\item {\em Size-constrained assignment matrix set:}
	For any $\bX \in [0,1]^{n \times r}$ with $\bX^\top \bone = \bkappa$ and $\bX \bone \leq \bone$, we have the error bounds
	\begin{subequations} \label{eq:error_Unrk}
		\begin{align}
			\dist(\bX,\setU^{n,r}_\bka) & \leq 3 \sqrt{\bone^\top \bka} 
			\left[ \sum_{j=1}^r (\ka_j - s_{\ka_j}(\bx_j) ) \right]  \\
			& \leq 3 \sqrt{\bone^\top \bka} ( \bone^\top \bka - \| \bX \|_{\rm F}^2 ).
		\end{align}
	\end{subequations}

	\item {\em Non-negative semi-orthogonal matrix set:}
	Recall that the expression of the convex hull of $\setS^{n,r}_+$ is not known.
	Let us consider $\setB^{n,r}_+:= \setB^{n,r} \cap \Rbb_+^{n \times r}$.
	For any $\bX \in \setB^{n,r}_+$, 
	we have the error bound
	\begin{align} \label{eq:error_Splus}
	\dist(\bX,\setS^{n,r}_+) \leq 5 r^{\frac{3}{4}} \sqrt{ r  - \| \bX \|_{\rm F}^2}.
	\end{align}

    \item {\em Cartesian product of CM sets:}
    Let $\bx = (\bx_1,\ldots,\bx_r)$, where $\bx_i \in \Rbb^{n_i}$.
    It is easy to show that 
    \[
    \dist(\bx,\conv(\setV)) \leq \textstyle \sum_{i=1}^r \dist(\bx_i,\conv(\setV_i)).
    \]
    If every $\setV_i$ has an error bound $\dist(\bx_i,\conv(\setV_i)) \leq \nu ( C_i - \| \bx_i \|_2^2 )$ for some $\nu > 0$ and for any $\bx_i \in \conv(\setV_i)$, then we have an error bound $\dist(\bx,\conv(\setV)) \leq \nu ( \sum_{i=1}^r C_i - \| \bx \|_2^2 )$ for any $\bx \in \conv(\setV)$.
 
\end{enumerate}
Applying the above error bound results to Lemmas \ref{lem:ep2}--\ref{lem:ep3} gives us the following important conclusion.
\begin{Theorem} \label{thm:err_bnd_spec_result}
	Consider the CM problem in \eqref{eq:cm}.
	Suppose that Assumption~\ref{asm:Lip2} holds.
	\begin{enumerate}[(a)]
		\item 
		Suppose that $\setV$ is any one of the above studied sets, except for the non-negative semi-orthogonal matrix set.		
		Given any scalar 
        $\lambda > K \nu$
        for some constant $\nu > 0$, 
		the EXPP problem in \eqref{eq:expp} or in \eqref{eq:expp_restate} is an equivalent formulation of the CM problem \eqref{eq:cm} in 
		the sense that the optimal solution sets of the CM problem and the EXPP problem are equal.
		
		\item 	
		Suppose that $\setV$ is the non-negative semi-orthogonal matrix set $\setS_+^{n,r}$.
		If we modify the EXPP problem by replacing the constraint $\bX \in \conv(\setS^{n,r}_+)$ with $\bX \in \setB^{n,r}_+$, then the EXPP problem is an inexact formulation of the CM problem in the sense of the assertions in Lemma~\ref{lem:ep3}.
		
	\end{enumerate}
\end{Theorem}

We should remind the reader that the assertions in Lemma~\ref{lem:ep3} imply that the EXPP problem for Case (b) becomes closer to the CM problem as $\lambda$ increases.
In the next subsection we will describe how the above error bounds for the various CM sets are shown. 

\subsection{How the CM Set Error Bounds are Shown}

Here we show the error bounds of the various CM sets in the previous subsection.

\begin{enumerate}[1.]
	\item {\em Binary vector set:} 
	It was shown in Section \ref{sect:ep2:1b_mimo_precode}.
	
	\item {\em MPSK set:} 
	The proof is relegated to {\revcolor Appendix \ref{appendix:MPSK_eb}.}
	
	\item {\em Unit sphere:} 
	For $\bx = \bzero$, Eq.~\eqref{eq:err_unit_sphere} holds trivially.
For $\bx \neq \bzero$, it can be verified that $\Pi_{\setS^n}(\bx) = \bx / \| \bx \|_2$.
Hence, for any $\| \bx \|_2 \leq 1$, $\bx \neq \bzero$, we have 
\begin{align*}
	 \revcolor{\dist(\bx,\setS^n)} & {\revcolor= \| \bx - \bx/ \| \bx \|_2 \|_2 = | 1 - \| \bx \|_2 | }\\
	& {\revcolor= 1 - \| \bx \|_2 \leq 1 - \| \bx \|_2^2.}
\end{align*}
\item {\em Semi-orthogonal matrix set:} 
	Let $\bX \in \Rbb^{n \times r}$ with $\bsig(\bX) \leq \bone$.
    Let $\bX = \bU \bSig \bV^\top$ be the SVD of $\bX$,
    where $\bU \in \setS^{n,r}$, $\bV \in \setS^{r,r}$, and $\bSig = \Diag(\bsig(\bX))$.
    Let $\bY = \bU \bV^\top$. It can be shown that $\bY = \Pi_{\setS^{n,r}}(\bX)$, e.g., by the von Neumann trace inequality.
We have 
\begin{align*}
\dist(\bX,\setS^{n,r}) 
& =  \| \bX - \bU \bV^\top \|_{\fro} \\
& = \| \bone - \bsig(\bX) \|_2 
\leq  \| \bone - \bsig(\bX) \|_1 \\
& = \textstyle \sum_{i=1}^r (1 - \sigma_i(\bX) ) = r - \| \bX \|_* \\
& \leq \textstyle \sum_{i=1}^r (1 - \sigma_i(\bX)^2 ) =r - \| \bX \|_{\rm F}^2.
\end{align*}
	
	\item {\em Unit vector set:}
    Let $\bx \in [0,1]^n$ with $\bone^\top \bx = 1$.
Without loss of generality, assume that $x_1 \geq \cdots \geq x_n$.
It can be verified that $\| \bx - \be_1 \|_2 = \dist(\bx,\setU^{n})$.
We have 
\begin{align}
	\revcolor{\dist(\bx,\setU^{n})} & {\revcolor= \| \bx - \be_1 \|_2  \leq \| \bx - \be_1 \|_1 \nonumber} \\
	& {\revcolor= \textstyle (1-x_1) + \sum_{i= 2}^n x_i \nonumber}  \\
	& {\revcolor= \textstyle 2( 1 - x_1 );}
	\label{eq:err_Uk_proof10}
\end{align}
note that the second and third equalities are due to $0 \leq x_i \leq 1$ and $\bx^\top \bone = 1$, respectively.
Eq.~\eqref{eq:err_Uk_proof10} leads to the first error bound in \eqref{eq:err_U}.
Consider the following inequality:
\begin{align} \label{eq:err_Uk_proof11}
	x_1 & = x_1 (x_1 + x_2 + \cdots + x_n)  \geq x_1^2 + x_2^2 + \cdots + x_n^2.
\end{align}
Applying \eqref{eq:err_Uk_proof11} to \eqref{eq:err_Uk_proof10} gives the second error bound in \eqref{eq:err_U}.
	
	\item {\em Selection vector set:} 
    The proof method is the same as the previous.
    Let $\bx \in [0,1]^n$ with $\bone^\top \bx = \kappa$, and assume that $x_1 \geq \cdots \geq x_n$.
    Let $\by = (1,\ldots,1,0,\ldots,0)$, where the number of ones is $\ka$.
    It can be verified that $\| \bx - \by \|_2 = \dist(\bx,\setU_\ka^n)$.
    By the same spirit as \eqref{eq:err_Uk_proof10}, it can be shown that 
\begin{align}
	\dist(\bX,\setU^{n}_\kappa) & \leq \textstyle 2( \kappa - \sum_{i=1}^\kappa x_i ).
	\label{eq:err_Uk_proof1}
\end{align}
This results in the first error bound in \eqref{eq:err_Uk}.
The following lemma is a generalization of \eqref{eq:err_Uk_proof11}.
\begin{Lemma} \label{lem:for_Uk}
    For any $\bx \in [0,1]^n$ with $\bone^\top \bx= \kappa \in \{1,\ldots,n\}$ and $x_1 \geq \cdots \geq x_n$, it
    holds that $\sum_{i=1}^\kappa x_i \geq \| \bx \|_2^2$.
\end{Lemma}
The proof of Lemma \ref{lem:for_Uk} is given in Appendix \ref{app:lem:for_Uk}.
Applying Lemma \ref{lem:for_Uk} to \eqref{eq:err_Uk_proof1} gives the second error bound in \eqref{eq:err_Uk}.
	
	\item {\em Partial permutation matrix set:}
	The result is a special case of that of the size-constrained assignment matrix set.
		
	\item {\em Size-constrained assignment matrix set:}
	We relegate the proof to the next subsection.

	\item {\em Non-negative semi-orthogonal matrix set:}
	We need the following theorem.
\begin{Theorem}
	\label{thm:chen}
	{\bf (Theorem {\revcolor5} in \cite{chen2022tight})}
	For any $\bX \in \Rbb^{n \times r}$, we have the error bound
	\[
	\dist(\bX, \setS_+^{n,r}) \leq 5 r^{\frac{3}{4}} ( \| \bX_- \|_{\rm F}^\half + \| \bX^\top \bX - \bI \|_{\rm F}^\half ),
	\]
	where $\bX_- = \max\{ -\bX, \bzero \}$.
\end{Theorem}
We will discuss Theorem~\ref{thm:chen} in the second part of this paper. 
It can be verified that
$$\| \bX^\top \bX - \bI \|_{\rm F} = \| \bsig(\bX)^2 - \bone \|_2.$$
Hence, for any $\bX \in \Rbb^{n \times r}_+$ with $\bsig(\bX) \leq \bone$, we have
\begin{align*}
	{\revcolor\dist(\bX, \setS_+^{n,r})} & {\revcolor\leq 5 r^{\frac{3}{4}} \| \bsig(\bX)^2 - \bone \|_1^\half} \\ 
	& {\revcolor= 5 r^{\frac{3}{4}} \textstyle [ \sum_{i=1}^r ( 1 - \sigma_i(\bX)^2 ) ]^\half} \\
	& {\revcolor= 5 r^{\frac{3}{4}} ( r - \| \bX \|_{\rm F}^2 )^\half}.
\end{align*}
 
\end{enumerate}

\subsection{Proof of Error Bounds for Size-Constrained Assignment Matrices}
\label{sect:err_bnd_illus}

The proof of the error bounds for the size-constrained assignment matrix set $\setU_\bka^{n,r}$ is important as it will provide insight into our development in the second part of this paper.
In most of the above error bound proofs, 
we invariably do the following: i) given an $\bx \in \conv(\setV)$, find the projection $\by = \Pi_\setV(\bx)$; ii) analyze the error $\| \bx - \by \|_2$ and derive an error bound. This is possible when $\by$ is analytically tractable. The case here is not. We do not have a closed form for $\Pi_{\setU_\bka^{n,r}}(\bX)$.
Taking insight from the error bound analysis for non-negative semi-orthogonal matrices in \cite{jiang2023exact,chen2022tight}, we take the following approach.
We construct a $\bY$ by $\by_j = \Pi_{\setU_{\ka_j}^n}(\bx_j)$ for all $j$.
We take advantage of the fact that $\| \bx_j - \by_j \|_2$ is analytically tractable---we did so already for the case of $\setU_\ka^n$.
Taking the row constraint $\bX \bone \leq \bone$ into account, we aim to show that, when the errors $\| \bx_j - \by_j \|_2$'s are sufficiently small, $\bY$ will have to lie in $\setU_\bka^{n,r}$.
This will lead us to a local error bound, as we will see.
Once we have the local error bound, some simple trick can be used to establish a valid error bound.

The proof is as follows. Let $\bX \in [0,1]^{n \times r}$, with $\bX^\top \bone = \bkappa$ and $\bX \bone \leq \bone$, be given.
Let 
\[
h(\bX) = 2 \sum_{j=1}^r (\ka_j - s_{\ka_j}(\bx_j)).
\]
We consider two cases.
In the first case, we suppose that $$h(\bX) < 1.$$
Let $\bY \in \Rbb^{n \times r}$ be given by $\by_j = \Pi_{\setU_{\ka_j}^n}(\bx_j)$ for all $j$.
Using the first error bound for $\setU_\ka^n$ in
\eqref{eq:err_Uk},
we bound the error $\bX - \bY$ as
\ifconfver
\begin{align*}
{\revR1color \|\bX - \bY\|_{\fro} }&{\revcolor\leq\| \bX - \bY \|_{\ell_1} = 
\textstyle 
 \sum_{j=1}^r \| \bx_j - \by_j \|_1} 
\\
& 
{\revcolor\leq 2 
\textstyle  
\sum_{j=1}^r (\ka_j - s_{\ka_j}(\bx_j))} 
 \\
 &  
{\revcolor= h(\bX) < 1.}
\end{align*}
\else
\[
{\revcolor 
{\revR1color \|\bX - \bY\|_{\fro} \leq }
\| \bX - \bY \|_{\ell_1} = 
\textstyle 
 \sum_{j=1}^r \| \bx_j - \by_j \|_1 
\leq 2 
\textstyle  
\sum_{j=1}^r (\ka_j - s_{\ka_j}(\bx_j))   
= h(\bX) < 1.}
\]
\fi
Since $\bY$ is component-wise binary, it holds that  $\bone^\top \bar{\by}_i \in \{ 0, 1, 2, \ldots, r \}$ for all $i$.
If $\bone^\top \bar{\by}_i < 2$ for all $i$, then $\bY$ lies in $\setU_{\bka}^{n,r}$.
The former can be shown to be true under $h(\bX) < 1$:
    \ifconfver
\begin{align*}
{\revcolor\bone^\top \bar{\by}_i}  & {\revcolor= \bone^\top \bar{\bx}_i  + \bone^\top ( \bar{\by}_i - \bar{\bx}_i ) 
\leq 1 +  \| \bar{\by}_i - \bar{\bx}_i \|_1} \\
& 
{\revcolor\leq 1 + h(\bX) 
< 2},
\end{align*}
    \else
\[{\revcolor
\bone^\top \bar{\by}_i   = \bone^\top \bar{\bx}_i  + \bone^\top ( \bar{\by}_i - \bar{\bx}_i ) 
\leq 1 +  \| \bar{\by}_i - \bar{\bx}_i \|_1
\leq 1 + h(\bX) 
< 2,}
\]
    \fi 
where we have used $\bX \bone \leq \bone$, or $\bone^\top \bar{\bx}_i \leq 1$ for all $i$.
Since $\bY \in \setU_{\bka}^{n,r}$, we have
\beq \label{eq:localeb_Unrk}
\dist(\bX,\setU_{\bka}^{n,r}) \leq \| \bX - \bY \|_{\rm F} \leq  h(\bX).
\eeq 
This is the local error bound we mentioned previously.

In the second case we suppose that $$h(\bX) \geq 1.$$
We may not guarantee $\bY \in \setU_{\bka}^{n,r}$,
so we seek another direction.
The distance between $\bX$ and any $\bZ \in \setU_{\bka}^{n,r}$ is bounded, specifically,
\begin{align} \label{eq:BBB}
\| \bX - \bZ \|_{\rm F}^2 & = \| \bX \|_{\rm F}^2 - 2 \tr(\bX^\top \bZ) + \| \bZ \|_{\rm F}^2 
\leq 2 (\bone^\top \bka){\revcolor.} 
\end{align}
Here, we have $\tr(\bX^\top \bZ) \geq 0$ due to $\bX, \bZ \geq \bzero$,
and we have
$\ka_j = \sum_{i=1}^n x_{ij} \geq  \sum_{i=1}^n x_{ij}^2 = \| \bx_j \|_2^2$ 
(note that $0 \leq x_{ij} \leq 1$) and consequently $\| \bX \|_{\rm F}^2 \leq \bone^\top \bka$.
Let $B=\sqrt{2 (\bone^\top \bka)}$.
From \eqref{eq:BBB} we have
\begin{align} \label{eq:case2eb_Unrk}
\dist(\bX,\setU_{\bka}^{n,r}) & \leq 
 B \leq 
B \,
h(\bX).
\end{align}
Assembling the two cases, \eqref{eq:localeb_Unrk} and \eqref{eq:case2eb_Unrk}, together gives the first error bound in \eqref{eq:error_Unrk}.
The second error bound in \eqref{eq:error_Unrk} is obtained by applying Lemma \ref{lem:for_Uk} to the first error bound in \eqref{eq:error_Unrk}.

\subsection{Further Remarks}

Now that we have obtained exact penalization formulations in the form of~\eqref{eq:gen_prob2_ep2} for a large class of CM problems, it is natural to ask how we can tackle them numerically. If the objective function in question has a Lipschitz continuous gradient over $\setC$, then we can apply the vanilla projected gradient method as discussed in Section~\ref{subsec:pgm} to tackle the resulting formulation. On the other hand, if the objective function is weakly convex (but possibly non-smooth) on $\setC$, then it is still possible to tackle the resulting formulation using the vanilla projected subgradient method; see \cite{LZZS23}.

{\revR1color 
\section{Locally Optimal Solution Correspondences}
\label{sect:locmin}

Thus far, we have focused on exact penalization in the globally optimal sense; we identified conditions under which the globally optimal solution set of the EXPP problem equals that of the CM problem.
Using results in concave minimization and error bounds, we can also pin down correspondences between the locally optimal solution sets of the two problems.

\begin{Theorem} \label{thm:locmin}
    Consider the CM problem \eqref{eq:cm} and the EXPP problem \eqref{eq:expp}.
    Suppose that the CM set $\setV$ satisfies the condition that, for some constant $\nu > 0$, $\nu(C- \| \bx \|_2^2)$ is an error bound function of $\setV$ relative to $\conv(\setV)$.

    \begin{enumerate}[(a)]
        \item 
        Suppose that Assumption~\ref{asm:Lip}
        holds. 
        Given any scalar $\lambda > \max\{ L/2, K \nu \}$, the set of locally optimal solutions to the EXPP problem equals that to the CM problem.

        \item Suppose that Assumption~\ref{asm:Lip2}
        holds. 
        Given any scalar $\lambda > K \nu$,
        a locally optimal solution to the CM problem is a locally optimal solution to the EXPP problem. 
        Moreover, if a locally optimal solution to the EXPP problem lies in $\setV$, then it is a locally optimal solution to the CM problem.
    \end{enumerate}
\end{Theorem}
Here is a simplified interpretation of Theorem~\ref{thm:locmin}:
If the objective function is differentiable, then the locally optimal solution set of the EXPP problem should equal that of the CM problem.
If the objective function is nonsmooth, then any locally optimal solution to the CM problem should be a locally optimal solution to the EXPP problem; the converse, however, requires some additional condition.

The proof of Theorem~\ref{thm:locmin} is described as follows. 
We use the notation ${\rm arglocmin}_{\bx \in \setX} \, \phi(\bx)$ to denote the set of locally optimal solutions to the problem $\min_{\bx \in \setX} \phi(\bx)$.
Theorem~\ref{thm:locmin} is built on the following two results.
\begin{Lemma}{\bf (modified version of Theorem 2 in \cite{shao2019framework})} \label{lem:locmin_cm}
Consider the setting of the concave minimization principle in Lemma \ref{lem:cvx_min}.
Recall that $\setD \subseteq  \Rbb^n$,
$\setA \subseteq \setD$ is non-empty with $\conv(\setA) \subseteq \setD$,
and $\Phi: \setD \rightarrow \Rbb$ is strictly concave on $\conv(\setA)$.
It holds that
\[
\check{\bx} \in \mathop{\rm arglocmin}_{\bx \in \conv(\setA)} \, \Phi(\bx)
\quad 
\Longrightarrow
\quad 
\check{\bx} \in \mathop{\rm arglocmin}_{\bx \in \setA} \, \Phi(\bx)
\]
\end{Lemma}

\begin{Lemma}{\bf (Proposition 9.1.2 in \cite{CP22})} \label{lem:locmin_eb}
    Consider the setting of the error bound principle in Lemma \ref{lem:ep2}.
    Recall that $\setD \subseteq  \Rbb^n$,
    $\setA \subseteq \setD$ is non-empty closed,
    $\setC \subseteq \setD$ is some set such that $\setA \subseteq \setC$,
    $\phi: \setD \rightarrow \Rbb$ is $K$-Lipschitz continuous on $\setC$,
    and $\psi: \setD \rightarrow \Rbb$ is an error bound function of $\setA$ relative to $\setC$.
\begin{enumerate}[(a)]
    \item Given any scalar $\lambda > K$, it holds that 
    \[
\check{\bx} \in \mathop{\rm arglocmin}_{\bx \in \setA} \, \phi(\bx)
~ 
\Longrightarrow
~ 
\check{\bx} \in \mathop{\rm arglocmin}_{\bx \in \setC} \, \phi(\bx) + \lambda \psi(\bx).
\]
    \item 
    If $\check{\bx}$ is a point in $\setA$, then the following implication is true:
    \[
    \check{\bx} \in \mathop{\rm arglocmin}_{\bx \in \setC} \, \phi(\bx) + \lambda \psi(\bx)
~ 
\Longrightarrow
~ 
\check{\bx} \in \mathop{\rm arglocmin}_{\bx \in \setA} \, \phi(\bx).
\]
\end{enumerate}
\end{Lemma}
We provide the proofs of Lemmas~\ref{lem:locmin_cm} and \ref{lem:locmin_eb} in  Appendices \ref{appendix:proof:lem:locmin_cm} and \ref{appendix:lem:locmin_eb}, respectively, for readers' easy access and for clear exposition of the underlying ideas.
Let us first consider Theorem \ref{thm:locmin}(a).
Applying Lemma~\ref{lem:locmin_cm} (and also Lemma~\ref{lem:cvx_push}) to the EXPP problem gives the result that a locally optimal solution to the EXPP problem is that to the CM problem.
Assumption~\ref{asm:Lip} implies Assumption~\ref{asm:Lip2}, and consequently we can apply Lemma~\ref{lem:locmin_eb}(a) to obtain the result that a locally optimal solution to the CM problem is that to the EXPP problem.
The proof of Theorem \ref{thm:locmin}(a) is therefore complete.
Moreover, Theorem \ref{thm:locmin}(b) is the direct corollary of Lemmas~\ref{lem:locmin_eb}(a)--(b).

}

\section{Conclusion}
\label{sect:conclusion}

We developed a framework for a class of CM problems.
Called EXPP, this framework converts a CM problem to a convex-constrained penalized formulation that has benign structures from the viewpoint of building algorithms.
In particular, we can handle the penalized formulation by using methods as simple as the vanilla projected gradient or subgradient method. 
A central aspect is whether EXPP is an exact formulation of a CM problem.
Also, the computation of the projected gradient method for each type of CM constraint is of concern.
The following is a summary of our exploration.
\begin{enumerate}[1.]
    \item 
    Assume that the objective function $f$ satisfies the Lipschitz continuous gradient assumption in Assumption~\ref{asm:Lip}; any twice {\revcolor continuously} differentiable $f$ falls into this scope.
    EXPP is an exact penalization formulation for {\em any} CM constraint set.
    This is the outcome under the concave minimization principle.
    One can straightforwardly apply the projected gradient method to implement EXPP.

    \item Assume that $f$ satisfies the Lipschitz continuous assumption in Assumption~\ref{asm:Lip2};
    any {\revcolor continuously} differentiable $f$ or any non-differentiable weakly-convex $f$ on an open domain (say, $\Rbb^n$) falls into this scope.
    EXPP is {\em not} guaranteed to be an exact penalization formulation for {\em any} CM set;
    it can at best be an inexact penalization formulation whose exactness improves as the penalty parameter $\lambda$ increases.
    However, EXPP is an exact penalization formulation for many CM sets of interest.
    These are the outcomes under the error bound principle.
    We may apply the projected gradient or subgradient method to implement EXPP.

    \item 
    When we work on a specific CM set $\setV$, we need to exploit the underlying structure.
    The computational efficiency of the projected gradient or subgradient method depends on whether the projection onto $\conv(\setV)$ can be efficiently computed.
    We saw easy cases, and we also saw challenges.
    
\end{enumerate}

In Part II of this paper, we will provide numerical results on different applications. We will also continue our study for some challenging CM cases.
As a remark, while we focus on the projected gradient method to demonstrate the utility of EXPP, it is free for one to consider other methods, such as the Frank-Wolfe method or the ADMM method, to implement EXPP.
The opportunity for such further development is open.

\ifplainver\section*{Appendix} \else \appendix \fi

\ifplainver\renewcommand\thesubsection{\Alph{subsection}} \renewcommand\thesubsubsection{\Alph{subsection}.\arabic{subsubsection}}\else \fi

\subsection{Proof of Lemma \ref{lem:ep3}}
\label{app:proof_lem:ep3}

Let $\Phi(\bx) = \phi(\bx) + \lambda \psi(\bx)^2$.
First we have 
\beq \label{eq:lem:ep3:proof_eq1}
\phi(\bx^\star) = \Phi(\bx^\star) \geq \Phi(\hat{\bx}) \geq \phi(\hat{\bx}).
\eeq 
Second,
\begin{subequations} \label{eq:lem:ep3:proof_eq2}
	\begin{align} 
		\label{eq:lem:ep3:proof_eq2a}
		\Phi(\hat{\bx}) & \geq 
			\phi(\bx') - K \| \hat{\bx} - \bx' \|_2 + \lambda \| \hat{\bx} - \bx' \|_2^2 \\
		\label{eq:lem:ep3:proof_eq2b}
			& \geq \Phi(\hat{\bx}) - K \| \hat{\bx} - \bx' \|_2 + \lambda \| \hat{\bx} - \bx' \|_2^2,
	\end{align}
\end{subequations}
where \eqref{eq:lem:ep3:proof_eq2a} is due to the Lipschitz continuity of $\phi$ and to the inequality $\psi(\hat{\bx})^2 \geq \dist(\hat{\bx},\setA)^2 = \| \hat{\bx} - \bx' \|_2^2$,
and \eqref{eq:lem:ep3:proof_eq2b} is due to $\phi(\bx') = \Phi(\bx') \geq \Phi(\hat{\bx})$.
Eq.~\eqref{eq:lem:ep3:proof_eq2} implies that 
\beq \label{eq:lem:ep3:proof_eq3} 
\frac{K}{\lambda} \geq \| \hat{\bx} - \bx' \|_2 = \dist(\hat{\bx},\setA).
\eeq 
Third we have 
\begin{align}
	\phi(\bx^\star) & \leq \phi(\bx') \leq \phi(\hat{\bx}) + K \| \hat{\bx} - \bx' \|_2  \leq \phi(\hat{\bx}) + K^2/\lambda,
	\label{eq:lem:ep3:proof_eq3_5}
\end{align}
where we have used \eqref{eq:lem:ep3:proof_eq3} and the Lipschitz continuity of $\phi$.
Fourth,
\begin{subequations} \label{eq:lem:ep3:proof_eq4}
	\begin{align}
		\label{eq:lem:ep3:proof_eq4a}
		\phi(\bx') & \leq \phi(\hat{\bx}) + K \| \hat{\bx} - \bx' \|_2 \\
		\label{eq:lem:ep3:proof_eq4b}
		& \leq \Phi(\hat{\bx}) + K^2/\lambda \\
		\label{eq:lem:ep3:proof_eq4c} 
		& \leq \phi(\bx^\star) + K^2/\lambda,
	\end{align}
\end{subequations}
where \eqref{eq:lem:ep3:proof_eq4a} is once again due to the Lipschitz continuity of $\phi$,
 \eqref{eq:lem:ep3:proof_eq4b} is due to \eqref{eq:lem:ep3:proof_eq3},
 and \eqref{eq:lem:ep3:proof_eq4c} is due to \eqref{eq:lem:ep3:proof_eq1}.
 Eqs. \eqref{eq:lem:ep3:proof_eq1}, \eqref{eq:lem:ep3:proof_eq3}, \eqref{eq:lem:ep3:proof_eq3_5} and  \eqref{eq:lem:ep3:proof_eq4} lead to the assertions in Lemma \ref{lem:ep3}.
  
 \subsection{Proof of Theorem \ref{thm:err_bnd_gen}(a)}
 \label{app:thm:err_bnd_gen:a}
 
As described in the theorem, we consider $\setV= \{ e^{-\jj \varphi}, e^{\jj \varphi} \}$, $\varphi \in (0,\pi/2]$, as a counter-example.
Any point $x$ in $\text{conv}(\mathcal{V})$ can be characterized as
\begin{align*}
x & = \theta e^{-\jj \varphi} + (1-\theta) e^{\jj \varphi} 
= \cos \varphi + \jj (1-2\theta) \sin \varphi,
\end{align*}
where $\theta \in [0,1]$.
Since $1-2\theta \in [-1,1]$, we can re-characterize $x$ as
\begin{align*}
x & = \cos \varphi + \jj \beta \sin \varphi,
\quad  \beta \in [-1,1].
\end{align*}
It can be verified that
\beq  \label{eq:thm:err_bnd_gen_a0}
\dist(x,\setV) = | x - e^{\jj {\rm sgn}(\beta) \varphi} | = ( 1- |\beta | ) \sin \varphi,
\eeq 
where ${\rm sgn}(\beta)= 1$ if $\beta \geq 0$ and ${\rm sgn}(\beta)= -1$ if $\beta < 0$.
From the absolute square $|x|^2 = \cos^2(\varphi) + \beta^2 \sin^2(\varphi)$ we have
\beq \label{eq:thm:err_bnd_gen_a1}
|\beta| \sin \varphi  = \sqrt{ |x|^2 - \cos^2(\varphi) }{\revcolor.}
\eeq 
As a basic inequality, it can be verified that
\[
\sqrt{a} \leq \sqrt{b} + \frac{1}{2\sqrt{b}}(a-b),
\quad \text{for any $a \geq 0,\  b > 0.$}
\]
Applying the above inequality to the right-hand side of \eqref{eq:thm:err_bnd_gen_a1}, with $a= |x|^2 - \cos^2(\varphi)$ and $b= 1  - \cos^2(\varphi) = \sin^2(\varphi)$, gives 
\[
|\beta| \sin \varphi  \leq \sin \varphi + 
\frac{|x|^2-1}{2 \sin\varphi}{\revcolor.} 
\]
Putting the above inequality into \eqref{eq:thm:err_bnd_gen_a0} gives the desired result
\[
\dist(x,\setV) \geq \frac{1- |x|^2}{2 \sin\varphi}. 
\]

{\revMovecolor\subsection{Proof of \eqref{eq:err_mpsk}, the Error Bound for the MPSK Set}
\label{appendix:MPSK_eb}

We need the following lemmas.
\begin{Lemma} \label{lem:mpsk1}
    Let $\varphi \in [0,2\pi)$.
    Let $y \in \Cbb$. It holds that
    $y \in \conv(\{0, e^{-j\varphi}, e^{j\varphi} \})$
    if and only if $y$ takes the form
    \beq \label{eq:lem:mpsk1}
    y = \alpha \left[ \cos\left( \varphi \right) + \jj \beta \sin\left( \varphi \right) \right],
    \eeq 
    for some $\alpha \in [0,1]$, $\beta \in [-1,1]$.
\end{Lemma}

\medskip
{\em Proof of Lemma~\ref{lem:mpsk1}:} \
Let  $y$ be any point in $\conv(\{0, e^{-j\varphi}, e^{j\varphi} \})$.
We can write $y = \theta_2 e^{-j\varphi} + \theta_3 e^{j\varphi}$ for some $\btheta \in \Rbb_+^3$, $\theta_1 + \theta_2 + \theta_3 =1$.
If $\theta_2 = \theta_3 = 0$, we can represent $y$ by \eqref{eq:lem:mpsk1} with $\alpha = \beta = 0$.
If either $\theta_2 > 0$ or $\theta_3 > 0$, we can do the following. 
Let $\alpha = \theta_2 + \theta_3$, and let $\vartheta = \theta_2/\alpha$.
We can write
\ifconfver
\begin{align}\label{eq:lem:mpsk1_p1}
    {\revMoveInsidecolor y} & {\revMoveInsidecolor= \alpha ( \vartheta e^{-j\varphi} + (1-\vartheta) e^{j\varphi}) \nonumber} \\
    & {\revMoveInsidecolor= \alpha \left[ \cos\left( \varphi \right) + \jj (1-2\vartheta) \sin\left( \varphi \right) \right].}
\end{align}
\else
\begin{equation}\label{eq:lem:mpsk1_p1}
        {\revMoveInsidecolor y  = \alpha ( \vartheta e^{-j\varphi} + (1-\vartheta) e^{j\varphi})
     = \alpha \left[ \cos\left( \varphi \right) + \jj (1-2\vartheta) \sin\left( \varphi \right) \right].}
\end{equation}
\fi
By noting that $1-2\vartheta \in [-1,1]$, we see from \eqref{eq:lem:mpsk1_p1} that $y$ takes the form in \eqref{eq:lem:mpsk1}.
Conversely, let $y$ be given by \eqref{eq:lem:mpsk1}.
By letting $\vartheta=(1-\beta)/2$, we can see from \eqref{eq:lem:mpsk1_p1} that $y$ is a convex combination of $0, e^{-j\varphi}, e^{j\varphi}$.
The proof is complete.
\hfill $\blacksquare$
\medskip

\begin{Lemma} \label{lem:MPSK_err_bd}
    Let $\varphi \in (0,\pi/2)$.
    Let $y \in \conv(\{ 0, e^{-\jj \varphi}, e^{\jj \varphi} \})$. It holds that
    \[
    \dist(y,\{ e^{-\jj \varphi}, e^{\jj \varphi} \} ) \leq \nu  (1-|y|^2),
    \]
    where $\nu = 1/\min\{ \sin(\varphi), \cos(\varphi) \}$.
\end{Lemma}

\medskip
{\em Proof of Lemma \ref{lem:MPSK_err_bd}:} \
Let $y \in \conv(\{ 0, e^{-\jj \varphi}, e^{\jj \varphi} \})$.
We can represent $y$ by \eqref{eq:lem:mpsk1} for some $\alpha \in [0,1]$, $\beta \in [-1,1]$.
It can be verified that $\dist(y,\{ e^{-\jj \varphi}, e^{\jj \varphi} \} ) = | y - e^{\jj \varphi} |$ if $\beta \geq 0$, and that $\dist(y,\{ e^{-\jj \varphi}, e^{\jj \varphi} \} ) = | y - e^{-\jj \varphi} |$ if $\beta \leq 0$.
This leads to
\ifconfver
    \begin{align}
        & \dist(y,\{ e^{-\jj \varphi}, e^{\jj \varphi} \} )  \nonumber \\ 
        & \qquad \qquad = \left[ (1-\alpha)^2 \cos(\varphi)^2 + (1-\alpha|\beta| )^2 \sin(\varphi)^2 \right]^\half \nonumber \\
        & \qquad \qquad \leq (1-\alpha) \cos(\varphi) + (1-\alpha|\beta| ) \sin(\varphi). 
        \label{eq:lem:MPSK_err_bd_p1}
    \end{align}
\else
    \begin{align}
        \dist(y,\{ e^{-\jj \varphi}, e^{\jj \varphi} \} ) & = \left[ (1-\alpha)^2 \cos(\varphi)^2 + (1-\alpha|\beta| )^2 \sin(\varphi)^2 \right]^\half \nonumber \\
        & \leq (1-\alpha) \cos(\varphi) + (1-\alpha|\beta| ) \sin(\varphi). 
        \label{eq:lem:MPSK_err_bd_p1}
    \end{align}
\fi 
Consider the case $\cos(\varphi) \geq \sin(\varphi)$.
From \eqref{eq:lem:MPSK_err_bd_p1} we have
\ifconfver
    \begin{align*}
        & \dist(y,\{ e^{-\jj \varphi}, e^{\jj \varphi} \} )   \\
        & \qquad \leq \frac{(1-\alpha) \sin(\varphi) \cos(\varphi) + (1-\alpha|\beta| ) \sin(\varphi)^2}{\sin(\varphi)} \\
        & \qquad \leq \frac{(1-\alpha) \cos(\varphi)^2 + (1-\alpha|\beta| ) \sin(\varphi)^2}{\sin(\varphi)} \\
        & \qquad \leq \frac{1 - [ \alpha^2 \cos(\varphi)^2 + (\alpha|\beta|)^2  \sin(\varphi)^2 ]}{\sin(\varphi)} \\
        & \qquad = \frac{1 - |y|^2}{\sin(\varphi)},
    \end{align*}
\else
    \begin{align*}
        \dist(y,\{ e^{-\jj \varphi}, e^{\jj \varphi} \} ) & \leq \frac{(1-\alpha) \sin(\varphi) \cos(\varphi) + (1-\alpha|\beta| ) \sin(\varphi)^2}{\sin(\varphi)} \\
        & \leq \frac{(1-\alpha) \cos(\varphi)^2 + (1-\alpha|\beta| ) \sin(\varphi)^2}{\sin(\varphi)} \\
        & \leq \frac{1 - [ \alpha^2 \cos(\varphi)^2 + (\alpha|\beta|)^2  \sin(\varphi)^2 ]}{\sin(\varphi)} \\
        & = \frac{1 - |y|^2}{\sin(\varphi)},
    \end{align*}
\fi
where the third inequality is due to $a^2 \leq a$ for $a \in [0,1]$.
Similarly, it can be shown that, for $\sin(\varphi) \geq \cos(\varphi)$, it holds that $\dist(y,\{ e^{-\jj \varphi}, e^{\jj \varphi} \} ) \leq (1 - |y|^2)/\cos(\varphi)$.
Combining the two cases leads to the desired result.
\hfill $\blacksquare$
\medskip

Now we show the error bound \eqref{eq:err_mpsk} for the MPSK set  $\Theta_m$.
Let $x$ be any point in $\setP_m$.
We can represent $x$ by 
\beq \label{eq:proof_tttt1}
x = y e^{\jj \frac{2\pi k}{m}},
\eeq 
where $k \in \{ 0,\ldots,m-1\}$ and $y \in \conv(\{ 0, e^{-\jj \frac{\pi}{m}}, e^{\jj \frac{\pi}{m}} \})$.
This representation is seen to be true from pictures; see Fig.~\ref{fig:mpsk-convexhull}.
\ifconfver
{\revMoveInsidecolor It can also be algebraically shown, and we refer the reader to Proposition \ref{prop:MPSK_cvx} in the supplemental material for details.}
\else
{\revMoveInsidecolor It can also be algebraically shown, and we refer the reader to Proposition \ref{prop:MPSK_cvx} in Appendix \ref{appendix:MPSK_convex_hull} for details.}
\fi
It can be verified from \eqref{eq:proof_tttt1} that $\Pi_{\Theta_m}(x) \in \{ e^{\jj \frac{2\pi k}{m} - \jj \frac{\pi}{m}}, e^{\jj \frac{2\pi k}{m} + \jj \frac{\pi}{m}} \}$.
We hence have
\begin{align*}
    {\revMoveInsidecolor\dist(x,\Theta_m)} & {\revMoveInsidecolor = \dist(x,\{ e^{\jj \frac{2\pi k}{m} - \jj \frac{\pi}{m}}, e^{\jj \frac{2\pi k}{m} + \jj \frac{\pi}{m}} \} )} \\
    & {\revMoveInsidecolor= \dist(y, \{ e^{-\jj \frac{\pi}{m}}, e^{\jj \frac{\pi}{m}} \})} \\
    & {\revMoveInsidecolor\leq \nu (1-|y|^2) = \nu (1-|x|^2), }
\end{align*}
where $\nu = 1/\min\{ \sin(\pi/m), \cos(\pi/m) \}$.
Here, the {\finalcolor inequality} is due to Lemma \ref{lem:MPSK_err_bd}. 
By noting that $\min\{ \sin(\pi/m), \cos(\pi/m) \} = \cos(\pi/3) = 0.5$ for $m= 3$, and $\min\{ \sin(\pi/m), \cos(\pi/m) \}  = \sin(\pi/m)$ for $m \geq 4$, we get the desired result.}

\subsection{Proof of Lemma~\ref{lem:for_Uk}}
\label{app:lem:for_Uk}

Recall that $1 \geq x_1 \geq \cdots \geq x_n \geq 0$ and $\sum_{i=1}^n x_i = \ka$.
Let us write 
\begin{equation}\label{eq:lem:for_Uk_t1}\begin{aligned}
      &{\revcolor\sum_{i=1}^\ka x_i  = \sum_{i=1}^{\ka-1} x_i + a - b,} \\
&{\revcolor a = x_\ka \left( \ka - \sum_{i=1}^{\ka-1} x_i \right),\  b  = x_\ka \left( (\ka - 1) - \sum_{i=1}^{\ka-1} x_i \right).}  
    \end{aligned}
\end{equation}
We have
\begin{align*}
{\revcolor a} & {\revcolor= x_\ka \left( \sum_{i=\ka}^n x_i \right)
\geq \sum_{i=\ka}^n x_i^2,} \\
{\revcolor b} & {\revcolor= x_\ka \left( \sum_{i=1}^{\ka-1}  (1 - x_i) \right) \leq \sum_{i=1}^{\ka-1} x_i (1 - x_i) = \sum_{i=1}^{\ka-1} x_i - \sum_{i=1}^{\ka-1} x_i^2.}
\end{align*}
Putting the above inequalities into \eqref{eq:lem:for_Uk_t1} gives the desired result $\sum_{i=1}^\ka x_i \geq \sum_{i=1}^n x_i^2$.

{\revR1color 
\subsection{Proof of Lemma~\ref{lem:locmin_cm}}
\label{appendix:proof:lem:locmin_cm}

Let $\check{\bx}$ be a locally optimal solution to the problem $\min_{\bx \in \conv(\setA)} \Phi(\bx)$.
First, we show that $\check{\bx} \in \setA$.
By the definition of locally optimal solutions, there exists a constant $\eps > 0$ such that 
\beq  \label{eq:lem:loc_concave:t1}
\Phi(\check{\bx}) \leq \Phi(\bx), \quad \forall \bx \in \conv(\setA) \cap \setN,
\eeq 
where 
$\setN = \{ \bx \in \Rbb^n \mid \| \bx - \check{\bx} \|_2 \leq \eps \}.$
Suppose that  $\check{\bx} \notin \setA$.
Then there exist two points $\bx', \bx'' \in \conv(\setA)$, $\bx' \neq \bx''$, and a scalar $\theta \in (0,1)$ such that 
\[
\check{\bx} = \theta \bx' + (1-\theta) \bx''.
\]
Let 
\beq \label{eq:lem:loc_concave:t1.5}
\begin{aligned}
    \tilde{\bx}' & = \check{\bx} - r(\bx'- \bx'') = (\theta - r) \bx' + (1-(\theta-r)) \bx'', \\
    \tilde{\bx}'' & = \check{\bx} + r(\bx'- \bx'') = (\theta + r) \bx' + (1-(\theta+r)) \bx'',
\end{aligned}
\eeq 
for some $0 < r \leq \min\{ \theta, 1-\theta, \eps/\| \bx' - \bx'' \|_2 \}$.
Note that 
\beq \label{eq:lem:loc_concave:t1.75}
\check{\bx} = \tfrac{1}{2} \tilde{\bx}' + \tfrac{1}{2} \tilde{\bx}''.
\eeq 
It is easy to verify that $\theta- r \in [0,1]$, $\theta + r \in [0,1]$, and $r \| \bx' - \bx'' \|_2 \leq \eps$.
Consequently we see from \eqref{eq:lem:loc_concave:t1.5} that $\tilde{\bx}',\tilde{\bx}'' \in \conv(\setA) \cap \setN$.
Also, applying Jensen's inequality to \eqref{eq:lem:loc_concave:t1.75} gives
\beq \label{eq:lem:loc_concave:t2}
\Phi(\check{\bx}) > \tfrac{1}{2} \Phi( \tilde{\bx}' ) +\tfrac{1}{2} \Phi( \tilde{\bx}' ) \geq \min\{  \Phi( \tilde{\bx}' ), \Phi( \tilde{\bx}'' ) \},
\eeq 
where the strict inequality is due to the strict concavity of $\Phi$.
Eq.~\eqref{eq:lem:loc_concave:t2} and $\tilde{\bx}',\tilde{\bx}'' \in \conv(\setA) \cap \setN$ violate the locally optimal condition in \eqref{eq:lem:loc_concave:t1}.
Hence, by contradiction, we must have $\check{\bx} \in \setA$.

Second, we show that $\check{\bx}$ is a locally optimal solution to the problem $\min_{\bx \in \setA} \Phi(\bx)$. 
Eq.~\eqref{eq:lem:loc_concave:t1} implies that $\Phi(\check{\bx}) \leq \Phi(\bx)$ for all $\bx \in \setA \cap \setN$.
Since $\check{\bx} \in \setA$, we have the desired result.
The proof is complete.

\subsection{Proof of Lemma~\ref{lem:locmin_eb}}
\label{appendix:lem:locmin_eb}

Let $\Phi(\bx) = \phi(\bx) + \lambda \psi(\bx)$.
First, we show the assertion in Lemma~\ref{lem:locmin_eb}(a).
Let $\check{\bx}$ be a locally optimal solution to $\min_{\bx \in \setA} \phi(\bx)$.
There exists a constant 
$\eps > 0$ such that 
\beq  \label{eq:lem:loc_eb:t1}
\phi(\check{\bx}) \leq \phi(\bx), \quad \forall \bx \in \setA \cap \setN,
\eeq 
where 
\beq \label{eq:lem:loc_eb:t2}
\setN = \{ \bx \in \Rbb^n \mid \| \bx - \check{\bx} \|_2 \leq \eps \}.
\eeq 
Let 
\[
\setN' = \{ \bx \in \Rbb^n \mid \| \bx - \check{\bx} \|_2 \leq \eps/2 \}.
\]
Let $\bx \in \setC \cap \setN'$,
and let $\ba \in \setA$ be a point such that $\| \bx - \ba \|_2 = \dist(\bx,\setA)$ (such a point exists since $\setA$ is non-empty closed). 
By noting that
\[
\| \bx - \ba \|_2 = \min_{\tilde{\bx} \in \setA } \| \bx  - \tilde{\bx} \|_2 \leq \| \bx - \check{\bx} \|_2,
\]
we have
\begin{align*}
    \| \check{\bx} - \ba \|_2 & \leq \| \check{\bx} - \bx \|_2 + \| \bx - \ba \|_2 \\
    & \leq 2 \| \check{\bx} - \bx \|_2 \leq \eps.
\end{align*}
This implies that $\ba \in \setN$.
Suppose that $\lambda > K$.
We have, for any $\bx \in \setC \cap \setN'$,
\begin{subequations} \label{eq:lem:loc_eb:t3}
\begin{align}
    \Phi(\bx) & \geq \phi(\bx) + \lambda \, \dist(\bx,\setA) \label{eq:lem:loc_eb:t3a} \\
    & \geq \phi(\ba) + (\lambda - K ) \| \bx - \ba \|_2  \label{eq:lem:loc_eb:t3b} \\
    & \geq \phi(\ba)  \label{eq:lem:loc_eb:t3c} \\
    & \geq \phi(\check{\bx}) = \Phi(\check{\bx}), \label{eq:lem:loc_eb:t3d}
\end{align}
\end{subequations}
where \eqref{eq:lem:loc_eb:t3b} is due to the Lipschitz continuity result $| \phi(\bx) - \phi(\ba) | \leq K \| \bx - \ba \|_2$;
\eqref{eq:lem:loc_eb:t3d} is due to $\ba \in \setA \cap \setN$ and \eqref{eq:lem:loc_eb:t1}.
Eq.~\eqref{eq:lem:loc_eb:t3} implies that $\check{\bx}$ is a locally optimal solution to $\min_{\bx \in \setC} \Phi(\bx)$.

Second, we show the assertion in Lemma~\ref{lem:locmin_eb}(b).
Let $\check{\bx}$ be a locally optimal solution to $\min_{\bx \in \setC} \Phi(\bx)$.
For some $\eps > 0$, we have 
\beq  \label{eq:lem:loc_eb:t4}
\Phi(\check{\bx}) \leq \Phi(\bx), \quad  \forall \bx \in \setC \cap \setN,
\eeq 
where $\setN$ is given by \eqref{eq:lem:loc_eb:t2}.
Since $\setA \subseteq \setC$ and $\psi(\bx) = 0$ for any $\bx \in \setA$, \eqref{eq:lem:loc_eb:t4} implies that $\Phi(\check{\bx}) \leq \phi(\bx)$ for all $\bx \in \setA \cap \setN$. 
If $\check{\bx}$ lies in $\setA$, then we further have $\phi(\check{\bx}) \leq \phi(\bx)$ for all $\bx \in \setA \cap \setN$, that is, 
$\check{\bx}$ is a locally optimal solution to $\min_{\bx \in \setA} \phi(\bx)$.
The proof is complete.

}

\ifconfver
\bibliographystyle{ieeetr}
\bibliography{refs,refs_JB_part1}
\fi

\ifconfver
\newpage
\begin{IEEEbiography}[{\includegraphics[width=1.2in,height=1.25in,clip,keepaspectratio]{./bios/Junbin Liu/ljb.png}}]{Junbin Liu} received the B.S. degree
from the South China University of Technology, Guangzhou, China, in 2017 and M.S. degree from the University of Chinese Academy of Sciences in 2020.
He is currently pursuing the Ph.D. degree
with the Department of Electronic Engineering, the Chinese University of Hong Kong, under the supervision of Professor Wing-Kin Ma.

His research interests encompass statistical signal processing methods, optimization theories, and their wide-ranging applications.
\end{IEEEbiography}

\begin{IEEEbiography}[{\includegraphics[width=1.2in,height=1.25in,clip,keepaspectratio]{./bios/Ya Liu/ly.jpg}}]{Ya Liu} received the B.S. degree from the University of Electronic Science and Technology of China (UESTC), Chengdu, China, in 2018.
She is currently pursuing a Ph.D. degree with the Department of Electronic Engineering at the Chinese University of Hong Kong (CUHK), under the supervision of Professor Wing-Kin Ma.

Her research focuses include matrix factorization, signal processing, and optimization, along with their diverse applications.
\end{IEEEbiography}

\begin{IEEEbiography}[{\includegraphics[width=1in,height=1.25in,clip,keepaspectratio]{./bios/Wing-Kin Ma/wingkinma.jpg}}]{Wing-Kin Ma}
 (Fellow, IEEE) is currently a Professor with the Department of Electronic Engineering, The
Chinese University of Hong Kong (CUHK), Hong Kong. His research interests include signal processing, optimization and communications, with recent focus on i) optimization and statistical aspects with structured matrix factorization, with application to remote sensing and data science; and ii) coarsely quantized MIMO transceiver designs.

Dr. Ma has rich experience in editorial service, such as an Associate Editor, and then later, a Senior Area Editor, and then from 2021 to 2023, the Editor-in-Chief of the
{\sc IEEE Transactions on Signal Processing},
and many others. He was a Tutorial Speaker in EUSIPCO 2011 and ICASSP 2014, and  was an IEEE Signal Processing Society (SPS) Distinguished Lecturer in 2018--2019. 	 
He was the recipient of the Research Excellence Award 2013--2014 by CUHK, the 2015 IEEE Signal Processing Magazine Best Paper Award, the 2016 IEEE Signal Processing
Letters Best Paper Award, and the 2018 IEEE SPS
Best Paper Award. He served as a Member of the Signal Processing
for Communications and Networking Technical Committee (SPCOM-TC) in 2015--2020,
a Member of 
Signal Processing Theory and Methods Technical Committee (SPTM-TC) in 2012--2017, 
the SPS Regional
Director-at-Large for Region 10 in 2020--2021, and a Technical Program Co-Chair of ICASSP 2023.
He co-founded and co-organized One World Signal Processing in 2020, a virtual seminar series for signal processing.
	
\end{IEEEbiography}

\begin{IEEEbiography}[{\includegraphics[width=1.2in,height=1.25in,clip,keepaspectratio]{./bios/Mingjie Shao/MingjieShao.jpg}}]{Mingjie Shao}
(S’16-M'20) received the B.S. degree
from the Xidian University, Xi'an, China, in 2015 and Ph.D. degree from the Chinese University of Hong Kong (CUHK) in 2020.
He was a Postdoctoral Fellow
with the Department of Electronic Engineering, CUHK from 2020 to 2023.
He is currently a  Research Professor (Qilu Young Scholar) in the School of Information Science and Engineering, Shandong University, Qingdao, China.
He was the recipient of the Hong Kong PhD Fellowship Scheme (HKPFS) from August 2015.
He was listed in the Student Best Paper Finalists in ICASSP 2017.

His research interests
focus on convex and non-convex
optimization, statistical signal processing and machine learning for wireless communication.

\end{IEEEbiography}

\begin{IEEEbiography}[{\includegraphics[width=1in,height=1.25in,clip,keepaspectratio]{bios/Anthony Man-Cho So/AnthonySo.jpg}}]{Anthony Man-Cho So} (Fellow, IEEE) received the B.S.E. degree in computer science from Princeton University, Princeton, NJ, USA, with minors in applied and computational mathematics, engineering and management systems, and German language and culture; the M.Sc. degree in computer science and the Ph.D. degree in computer science with a Ph.D. minor in mathematics from Stanford University, Stanford, CA, USA. He is currently Dean of the Graduate School, Deputy Master of Morningside College, and a Professor with the Department of Systems Engineering and Engineering Management at The Chinese University of Hong Kong (CUHK), Hong Kong SAR, China. His research interests include optimization theory and its applications in various areas of science and engineering, including computational geometry, machine learning, signal processing, and statistics. 
	
Dr. So is a Fellow of the Hong Kong Institution of Engineers. He is the recipient of a number of research and teaching awards, including the 2024 INFORMS Computing Society Prize, the SIAM Review SIGEST Award in 2024, the 2022 University Grants Committee Teaching Award, the 2018 IEEE Signal Processing Society Best Paper Award, the 2015 IEEE Signal Processing Society Signal Processing Magazine Best Paper Award, the 2014 IEEE Communications Society Asia-Pacific Outstanding Paper Award, the 2013 CUHK Vice-Chancellor’s Exemplary Teaching Award, and the 2010 INFORMS Optimization Society Optimization Prize for Young Researchers. He currently serves on the Editorial Boards of \emph{Journal of Global Optimization}, \emph{Mathematics of Operations Research}, \emph{Mathematical Programming}, \emph{Open Journal of Mathematical Optimization}, \emph{Optimization Methods and Software}, and \emph{SIAM Journal on Optimization}. He was also the Lead Guest Editor of the Special Issue on Non-Convex Optimization for Signal Processing and Machine Learning of the {\sc IEEE Signal Processing Magazine} and a Guest Editor of the Special Issue on Advanced Optimization Theory and Algorithms for Next Generation Wireless Communication Networks of the {\sc IEEE Journal on Selected Areas in Communications}.
\end{IEEEbiography}

\fi

\ifconfver
\newpage
\setcounter{page}{1}
\title{Supplemental Material of {``Extreme Point Pursuit---Part I: A Framework for Constant Modulus Optimization''}}
\author{Junbin Liu, Ya Liu, Wing-Kin Ma, Mingjie Shao and Anthony Man-Cho So}
\maketitle

\ifplainver\renewcommand\thesubsection{\arabic{section}.\arabic{subsection}} \else \renewcommand\thesection{\Roman{section}} \fi

\fi

\ifconfver
\section{Derivations of the Convex Hull and Projection Results}\label{appendix:cvx_hull&proj}
\else
\subsection{Derivations of the Convex Hull and Projection Results}\label{appendix:cvx_hull&proj}
\fi


The purpose of this section is to provide the proofs of some of the main convex hull and projection results in Section \ref{sect:cvx_hull_proj}.
We will consider (a) the semi-orthogonal matrix set $\setS^{n,r}$, (b) the size-constrained assignment matrix set $\setU_\bka^{n,r}$, and (c) the MPSK set $\Theta_m$.
Except for the non-negative semi-orthogonal matrix set $\setS_+^{n,r}$, the above three cases subsume the other considered CM sets.

Before we proceed, we should mention that 
(a) the results for $\setS^{n,r}$ are considered known, but it is helpful to provide the details in this
\ifconfver 
{\revcolor supplemental material;}
\else
{\revcolor Appendix;}
\fi
(b) for the case of $\setU_\bka^{n,r}$, we only found a reference  \cite{sidiropoulos2015signal} that loosely mentioned the insight of the convex hull proof for the  case of $\sum_{i=1}^r \ka_i = n$ (not for the general case of $\sum_{i=1}^r \ka_i \leq n$);
(c) an algebraic proof for $\Theta_m$ appears unavailable in the literature.

\ifconfver
\subsection{Semi-Orthogonal Matrix Set}
\else
\subsubsection{Semi-Orthogonal Matrix Set}
\fi

To show that the convex hull of the semi-orthogonal matrix set $\setS^{n,r}$ is the unit spectral norm ball $\mathcal{B}^{n,r}= \{ \bX \in \Rbb^{n \times r} \mid \sigma_1(\bX) \leq 1 \}$,
we first note that any point $\bV$ in $\setS^{n,r}$ has $\sigma_1(\bV)= 1$. 
Let $\bX$ be any convex combination of points in $\setS^{n,r}$, i.e.,
$\bX = \sum_{i=1}^k \theta_i \bV_i$ for some $\bV_1,\ldots,\bV_k \in \setS^{n,r}$, $\btheta \in \Rbb^k_+$, $\sum_{i=1}^k \theta_i = 1$, and $k$.
Since $\sigma_1(\bX)$
is convex, it holds that $\sigma_1(\bX) \leq \sum_{i=1}^k \theta_i \sigma_1(\bV_i) = 1$.
Hence, $\bX$ lies in $\setB^{n,r}$.
Conversely, let $\bX$ be any point in  $\setB^{n,r}$.
Let $\bX = \bU \bSig \bV^\top$ be the SVD of $\bX$, where $\bU \in \Rbb^{n \times r}$ is semi-orthogonal, $\bSig = \Diag(\bsig)$ has $\bzero \leq \bsig \leq \bone$,
and $\bV \in \Rbb^{r \times r}$ is orthogonal.
Since $[-1,1]^r$ is the convex hull of $\{ -1, 1 \}^r$, $\bsig$ can be represented by $\bsig = \sum_{i=1}^k \theta_i \bw_i$ for some $\bw_1,\ldots,\bw_k \in \{-1,1\}^r$, $\btheta \in \Rbb^k_+$, $\sum_{i=1}^k \theta_i = 1$, and $k$.
Applying this representation to the SVD of $\bX$ gives
$\bX = \sum_{i=1}^k \theta_i \bW_i$, where $\bW_i = \bU \bD_i \bV^\top$, $\bD_i = \Diag(\bw_i)$, lies in  $\setS^{n,r}$ as one can easily verify. 
Hence, $\bX$ lies in $\conv(\setS^{n,r})$.
The proof of $\conv(\setS^{n,r})= \setB^{n,r}$ is complete.

We also show the formula for the projection onto $\setB^{n,r}$.
Given $\bZ \in \Rbb^{n \times r}$,
let $\bZ = \bU \bSig \bV^\top$ be the SVD of $\bZ$, where $\bU \in \Rbb^{n \times r}$ is semi-orthogonal, $\bSig= \Diag(\bm \sigma(\bZ))$,
	and $\bV \in \Rbb^{r \times r}$ is orthogonal.
For any $\bX \in \Rbb^{n \times r}$ with $\sigma_1(\bX) \leq 1$, it holds that
\begin{align*}
    \| \bZ - \bX \|_\fro & \geq \| \bsig(\bZ) - \bsig(\bX) \|_2 
    \geq \| \bsig(\bZ) - \bd \|_2,
\end{align*}
where the first inequality is due to the von Neumann trace inequality; 
the vector $\bd \in [0,1]^r$ has each of its components given by $d_i = 1$ for $\sigma_i(\bZ) > 1$ and $d_i = \sigma_i(\bZ)$ for $\sigma_i(\bZ) \leq 1$.
Moreover, equality with the above equation is attained if $\bX = \bU \bD \bV^\top$, where $\bD = \Diag(\bd)$.
It follows that the projection of $\bZ$ onto $\setB^{n,r}$ is 
  $\Pi_{\setB^{n,r}}(\bZ)= \bU \Diag([ \bm \sigma(\bZ)]_{\bzero}^{\bone}) \bV^\top$.

\ifconfver
\subsection{Size-Constrained Assignment Matrix Set}
\else
\subsubsection{Size-Constrained Assignment Matrix Set}
\fi

Recall
\[
	\setU^{n,r}_{\bka} = \{ \bX \in \{ 0, 1 \}^{n \times r} \mid  \bX^\top \bone = \bka, \bX \bone \leq \bone \}.
\]
Let
\beq \label{eq:pf_svs_eq1_1}
\setC = \{ \bX \in [ 0, 1 ]^{n \times r} \mid  \bX^\top \bone = \bka, \bX \bone \leq \bone \},
\eeq 
and let $\setW$ be the set of extreme points of $\setC$.
To show that $\conv(\setU_\bka^{n,r}) = \setC$, we divide our proof into two steps.

{\em Step 1:} \ 
Suppose that $\setW \subseteq \setU_\bka^{n,r}$.
Since $\setC$ is convex and compact, the Krein-Milman theorem asserts that $\setC = \conv(\setW)$.
It follows that $\setC \subseteq \conv(\setU_\bka^{n,r})$.
Also, it can be easily verified that any convex combination of points in $\setU_\bka^{n,r}$ lies in $\setC$, and so $\conv(\setU_\bka^{n,r}) \subseteq \setC$.
Hence, under the assumption that $\setW \subseteq \setU_\bka^{n,r}$, we have $\conv(\setU_\bka^{n,r}) = \setC$.

{\em Step 2:} \ 
We show that $\setW \subseteq \setU_\bka^{n,r}$ is true.
To do this, we apply results from integer linear programming.
Let us provide the context.
Given $\bA \in \Rbb^{m \times n}$ and $\bb \in \Rbb^m$, consider a polyhedron
\beq \label{eq:pf_svs_eq2}
    \setP = \{ \bx \in \Rbb^n \mid \bA \bx \leq \bb \}. 
\eeq
It is known that 
any extreme point of $\setP$ lies in  $\Zbb^n$ if $\bA$ is totally unimodular (TU) and $\bb$ lies in $\Zbb^m$ \cite[Theorem 19.1]{schrijver1998theory}.
Specifically, a matrix $\bA$ is said to be TU if every square submatrix of $\bA$ has determinant equal to $0$, $-1$, or $1$.
The set $\setC$ in \eqref{eq:pf_svs_eq1_1} can be represented by \eqref{eq:pf_svs_eq2}, with 
\beq \label{eq:pf_svs_eq3}
\bA = \begin{bmatrix}
    -\bI_{nr} \\ \bI_r \otimes \bone_n^\top \\ \bone_r^\top \otimes \bI_n \\  -\bI_r \otimes \bone_n^\top 
\end{bmatrix},
\qquad \bb = \begin{bmatrix}
    \bzero \\ \bka \\ \bone_n \\  -\bka 
\end{bmatrix},
\eeq 
where $\bI_{n}$ denotes the $n \times n$ identity matrix; $\bone_n$ denotes the length-$n$ all-one vector;
$\otimes$ denotes the Kronecker product;
$\bx$ in \eqref{eq:pf_svs_eq2} is the vectorization of $\bX$.
We see that $\bb$ has integer components.
Suppose that $\bA$ is TU.
Then, by the above stated integer linear programming result, 
every $\bW \in \setW$ lies in $\Zbb^{n \times r}$.
This, together with $\setW \subseteq \setC$, imply that $\setW \subseteq \setU^{n,r}_{\bka}$.

The remaining task is to show that $\bA$ is TU.
Let
\[
\bA' = \begin{bmatrix}
    -\bI \\ \bB \\ -\bB 
\end{bmatrix},
\qquad
\bB =
\begin{bmatrix}
    \bI_r \otimes \bone_n^\top \\ \bone_r^\top \otimes \bI_n
\end{bmatrix}.
\]
As a basic fact, if $\bB$ is TU, then $\bA'$ is TU (this fact was mentioned, e.g., in Eq.~(3), Chapter 19, in \cite{schrijver1998theory}).
To show that $\bB$ is TU, 
consider the following lemma.
\begin{Lemma} {\bf (Simplified version of Example 3, Chapter 19, in \cite{schrijver1998theory})}
    \label{lem:pr_svs_2}
    Let $\bA \in \Rbb^{m \times n}$ be a matrix with $\{0,1\}$ components.
    Suppose that each column of $\bA$ has exactly two nonzero components.
    Then $\bA$ is TU if and only if the rows of $\bA$ can be split into two classes such that, for each column, the two nonzero components in the column are in different classes.
\end{Lemma}
It can be verified that $\bB$ satisfies the premise of Lemma~\ref{lem:pr_svs_2}, and hence $\bB$ is TU. 
It follows that $\bA'$ is TU.
The matrix $\bA$ in \eqref{eq:pf_svs_eq3} is a submatrix of $\bA'$ obtained by removing the last $n$ rows of $\bA'$.
Since any submatrix of a TU matrix is, by definition, TU, we conclude that $\bA$ is TU.
The proof is complete.

\ifconfver
\subsection{MPSK Set, Part 1: Basic Lemmas}
\else
\subsubsection{MPSK Set, Part 1: Basic Lemmas}
\fi

The MPSK case seems easy.
We can draw the  MPSK set $\Theta_m$ (see, e.g., Fig. \ref{fig:mpsk-convexhull} in the main paper) and use observation to construct the convex hull and projection formulae.
It turns out that an algebraic proof for the MPSK results is tedious. 
Here we give an algebraic proof.
We start with deriving some basic trigonometric lemmas.

\begin{Lemma} \label{lem:mpsk2}
    Let $\varphi \in (0,\pi/2)$. Let $y \in \Cbb$. It holds that
    $\angle y \in [-\varphi,\varphi]$
    if and only if $y$ takes the form 
    \beq \label{eq:lem:mpsk1_new}
    y = \alpha \left[ \cos\left( \varphi \right) + \jj \beta \sin\left( \varphi \right) \right],
    \eeq    
    for some $\alpha \geq 0$, $\beta \in [-1,1]$.
    In particular we have $\angle y = \tan^{-1}( \beta \tan(\varphi ))$ and $|y| = \alpha \, r(\angle y)$, where $r(\phi) = \cos(\varphi) [ 1 + \tan(\phi)^2 ]^\half$. 
\end{Lemma}

\medskip
{\em Proof of Lemma~\ref{lem:mpsk2}:} \
Let $y = a e^{\jj \phi}$, where $a = |y|$ and $\phi = \angle y$.
Suppose that $y$ takes the form in \eqref{eq:lem:mpsk1_new} for some $\alpha \geq 0$, $\beta \in [-1,1]$.
The case of $\alpha = 0$ is trivial, and we focus on $\alpha > 0$.
Applying basic trigonometric results to \eqref{eq:lem:mpsk1_new} gives
\begin{align*}    
   {\revcolor \phi }
    & {\revcolor= \tan^{-1} \left( \frac{\beta \sin(\varphi)}{\cos(\varphi)}   \right) = \tan^{-1}( \beta \tan(\varphi )),} \\
    {\revcolor a}
    & {\revcolor= \alpha [ \cos(\varphi)^2 + \beta^2 \sin(\varphi)^2 ]^\half }\\ 
    & {\revcolor= \alpha \cos(\varphi) \left[ 1 + \beta^2 \tan(\varphi)^2   \right]^\half }\\
    &{\revcolor = \alpha \cos(\varphi) \left[ 1 + \tan(\phi)^2   \right]^\half = \alpha \, r(\phi).}
\end{align*}
Note the basic facts that $\tan(\varphi)$ is positive and bounded for $\varphi \in (0,\pi/2)$,
$\tan^{-1}(\omega)$ is monotone increasing on $(-\pi/2,\pi/2)$, and $\tan(-\omega) = - \tan(\omega)$.
It follows that $\phi \leq \tan^{-1}( \tan(\varphi )) = \varphi$,
and similarly $\phi \geq -\varphi$.
This concludes that $\phi \in [-\varphi,\varphi]$.
Conversely, suppose that $y$ has $\phi \in [-\varphi,\varphi]$.
Let us choose $\alpha = a/r(\phi)$ and $\beta = \tan(\phi)/\tan(\varphi) \in [-1,1]$.
It can be verified that
\begin{align*}
    \cos\left( \varphi \right) + \jj \beta \sin\left( \varphi \right) & = r(\phi) e^{\jj \phi}. 
\end{align*}
Consequently, we can represent $y$ by \eqref{eq:lem:mpsk1_new}.
The proof is complete.
\hfill $\blacksquare$
\medskip

\begin{Lemma}
	\label{lem:mpsk3}
	Let $m \geq 2$ be an integer.
    Let $\phi \in [-\pi/m,\pi/m]$.
	It holds that 
	\beq \label{eq:lem:mpsk3}
    \cos( \phi) 
    \geq \cos\left( \tfrac{2\pi l}{m} + 
    \phi
    \right),
	~ \forall l \in \{0,\ldots,m-1\}.
	\eeq 
\end{Lemma}

\medskip
{\em Proof of Lemma~\ref{lem:mpsk3}:} \
Let $\Omega = \{ 2 \pi l/m + \phi \mid l \in  \{0,\ldots,m-1\} \}$.
It is known that $\cos(\omega)$ is decreasing in $(0,\pi]$, 
that $\cos(\omega)$ is non-positive in $[\pi,3\pi/2)$,
and that $\cos(\omega)$ is increasing in $[3\pi/2,2\pi)$.
Suppose that $\phi \in [0,\pi/m]$.
Then
\ifconfver
\begin{align*}
\max_{ \omega \in \Omega \cap (0,\pi] } \cos(\omega) & = \cos\left( \phi \right), \\
\max_{ \omega \in \Omega \cap[\pi,3\pi/2) } \cos(\omega)
 & \leq 0, \\
\max_{ \omega \in \Omega \cap [3\pi/2,2\pi) } \cos(\omega)  & = \cos\left( \tfrac{2\pi (m-1)}{m} + \phi \right) 
= \cos\left( \tfrac{2\pi}{m} - \phi \right) \\
& \leq \cos\left( \phi \right),
\end{align*}
\else
$$
\begin{aligned}
 \max_{ \omega \in \Omega \cap (0,\pi] } \cos(\omega) & = \cos\left( \phi \right), \\
\max_{ \omega \in \Omega \cap[\pi,3\pi/2) } \cos(\omega)
 & \leq 0, \\
\max_{ \omega \in \Omega \cap [3\pi/2,2\pi) } \cos(\omega)  & = \cos\left( \tfrac{2\pi (m-1)}{m} + \phi \right) 
= \cos\left( \tfrac{2\pi}{m} - \phi \right) \leq \cos\left( \phi \right),
\end{aligned}
$$
\fi
and it follows that \eqref{eq:lem:mpsk3} holds.
Similarly we can show that \eqref{eq:lem:mpsk3} also holds for  $\phi \in [-\pi/m,0]$.
The proof is complete.
\hfill $\blacksquare$
\medskip

\medskip

\begin{Lemma}
	\label{lem:mpsk4}
	Let $m \geq 3$ be an integer.
	Let $y \in \Cbb$ be a variable with $\angle y \in [-\pi/m,\pi/m]$. 
	The variable $y$ satisfies 
	\beq \label{eq:lem:mpsk3:1}
		\Re \left(e^{\jj \frac{2\pi l}{m}} y \right) \leq \cos\left( \tfrac{\pi}{m} \right),
 		~ \forall l \in \{0,\ldots,m-1\}
	\eeq 
	if and only if $y$ takes the form in \eqref{eq:lem:mpsk1_new} 
	for some $\alpha \in [0,1]$, $\beta \in [-1,1]$.
\end{Lemma}

\medskip
{\em Proof of Lemma~\ref{lem:mpsk4}:} \
First suppose that \eqref{eq:lem:mpsk3:1} holds.
By Lemma~\ref{lem:mpsk2}, we can characterize $y$ as \eqref{eq:lem:mpsk1_new}  for some $\alpha \geq 0$ and $\beta \in [-1,1]$.
Putting \eqref{eq:lem:mpsk1_new} into \eqref{eq:lem:mpsk3:1} for $l= 0$ gives
\[
\cos\left( \tfrac{\pi}{m} \right) \geq \Re \left( y \right) = \alpha \cos\left( \tfrac{\pi}{m} \right).
\]
The above equation implies that $\alpha \leq 1$.
It follows that \eqref{eq:lem:mpsk1_new} holds for some $\alpha \in [0,1]$, $\beta \in [-1,1]$.

Second suppose that $y$ takes the form in \eqref{eq:lem:mpsk1_new} for some $\alpha \in [0,1]$, $\beta \in [-1,1]$.
The left-hand side of \eqref{eq:lem:mpsk3:1} can be written as
\beq \label{eq:lem:mpsk3:t1}
\Re \left(e^{\jj \frac{2\pi l}{m}} y \right)
= \alpha 
\left[ 
\cos\left( \tfrac{2\pi l}{m} \right) \cos\left( \tfrac{\pi}{m} \right) 
- \beta \sin\left( \tfrac{2\pi l}{m} \right) \sin\left( \tfrac{\pi}{m} \right)
\right]. 
\eeq 
For any $l \in \{0,\ldots,m-1\}$ satisfying $\sin(2\pi l/m) \geq 0$, we can bound \eqref{eq:lem:mpsk3:t1} as
\ifconfver
\begin{align*}
{\revcolor\Re \left(e^{\jj \frac{2\pi l}{m}} y \right)}
& {\revcolor\leq 
\cos\left( \tfrac{2\pi l}{m} \right) \cos\left( \tfrac{\pi}{m} \right) 
+ \sin\left( \tfrac{2\pi l}{m} \right) \sin\left( \tfrac{\pi}{m} \right)}
\\
& {\revcolor= \cos\left( \tfrac{2\pi l}{m} - \tfrac{\pi}{m} \right)} \\
& {\revcolor\leq \cos\left( \tfrac{\pi}{m} \right),}
\end{align*}
\else
 \[
{\revcolor \Re \left(e^{\jj \frac{2\pi l}{m}} y \right)
 \leq 
\cos\left( \tfrac{2\pi l}{m} \right) \cos\left( \tfrac{\pi}{m} \right) 
+ \sin\left( \tfrac{2\pi l}{m} \right) \sin\left( \tfrac{\pi}{m} \right)}
 {\revcolor= \cos\left( \tfrac{2\pi l}{m} - \tfrac{\pi}{m} \right) }
{\revcolor \leq \cos\left( \tfrac{\pi}{m} \right),}
\]
\fi
where the second inequality is due to Lemma~\ref{lem:mpsk3}.
Similarly, for any $l \in \{0,\ldots,m-1\}$ satisfying $\sin(2\pi l/m) \leq 0$, we have
\ifconfver
\begin{align*}
{\revcolor \Re \left(e^{\jj \frac{2\pi l}{m}} y \right)}
& {\revcolor \leq 
\cos\left( \tfrac{2\pi l}{m} \right) \cos\left( \tfrac{\pi}{m} \right) 
- \sin\left( \tfrac{2\pi l}{m} \right) \sin\left( \tfrac{\pi}{m} \right)} 
\\
& {\revcolor = \cos\left( \tfrac{2\pi l}{m} + \tfrac{\pi}{m} \right)}
\\
& {\revcolor \leq \cos\left( \tfrac{\pi}{m} \right).}
\end{align*}
\else
$$
{\revcolor \Re \left(e^{\jj \frac{2\pi l}{m}} y \right)
 \leq 
\cos\left( \tfrac{2\pi l}{m} \right) \cos\left( \tfrac{\pi}{m} \right) 
- \sin\left( \tfrac{2\pi l}{m} \right) \sin\left( \tfrac{\pi}{m} \right) 
 = \cos\left( \tfrac{2\pi l}{m} + \tfrac{\pi}{m} \right)
 \leq \cos\left( \tfrac{\pi}{m} \right).}
$$
\fi
It follows that \eqref{eq:lem:mpsk3:1} holds.
The proof is complete.
\hfill $\blacksquare$
\medskip

\ifconfver
\subsection{MPSK Set, Part 2: Convex Hull}
\label{appendix:MPSK_convex_hull}
\else
\subsubsection{MPSK Set, Part 2: Convex Hull}
\label{appendix:MPSK_convex_hull}
\fi

We now prove the convex hull of the MPSK set $\Theta_m$.
Recall 
$$\Theta_m= \{ x \in \Cbb \mid x= e^{\jj \frac{2\pi l}{m} + \jj \frac{\pi}{m} }, ~ l \in \{0,1,\ldots,m-1\} \},$$
where $m \geq 3$ is an integer.
Also, let 
    \ifconfver
        \begin{equation} \label{eq:Pm}
        \begin{aligned}
        \setP_m  =  \Big\{ x \in \Cbb & \, \Big| \, \Re \left(e^{\jj \frac{2\pi l}{m}} x \right) \leq \cos\left( \tfrac{\pi}{m} \right), \\
        & \qquad l \in \{0,\ldots,m-1\} \Big\}.
        \end{aligned}
	\end{equation}
    \else
     \begin{equation} \label{eq:Pm}
        \setP_m =  \big\{ x \in \Cbb \, \big| \, \Re \left(e^{\jj \frac{2\pi l}{m}} x \right) \leq \cos\left( \tfrac{\pi}{m} \right), ~ l \in \{0,\ldots,m-1\} \big\}. 
	\end{equation}
    \fi 
To show $\conv(\Theta_m)= \setP_m$,
we first show the following result.
\begin{Prop} \label{prop:Pm}
    Given any integer $m \geq 3$,  $\setP_m$ has an equivalent representation 
    \begin{align} \label{eq:Pm2}
        \setP_m & = \left\{ x = y e^{\jj \frac{2\pi k}{m}} \mid y \in \setT_m, k \in \{0,\ldots,m-1\} \right\},    \\
        \setT_m & = \conv(\{ 0, e^{-\jj \frac{\pi}{m}}, e^{\jj \frac{\pi}{m}} \}).
    \end{align} 
\end{Prop}

\medskip
{\em Proof of Proposition \ref{prop:Pm}:} \
Let $x$ be any point in $\Cbb$. We can characterize $x$ as $x = y e^{\jj \frac{2\pi k}{m}}$ for some $k \in \{0,\ldots,m-1\}$ and for some $y \in \Cbb$ with $\angle y \in [-\pi/m,\pi/m]$.
Consider the following condition
\[
\Re \left(e^{\jj \frac{2\pi l}{m}} x \right) \leq \cos\left( \tfrac{\pi}{m} \right), ~ \forall l \in \{0,\ldots,m-1\}. 
\]
The above condition is  equivalent to
\beq \label{eq:prop:Pm_p1}
\Re \left(e^{\jj \frac{2\pi l}{m}} y \right) \leq \cos\left( \tfrac{\pi}{m} \right), ~ \forall l \in \{0,\ldots,m-1\}. 
\eeq 
By Lemma~\ref{lem:mpsk4} and Lemma~\ref{lem:mpsk1}, \eqref{eq:prop:Pm_p1} holds if and only if $y \in \conv(\{ 0, e^{-\jj \frac{\pi}{m}}, e^{\jj \frac{\pi}{m}} \}) = \setT_m$.
It follows that the set $\setP_m$ in \eqref{eq:Pm} can be equivalently represented by \eqref{eq:Pm2}.
\hfill $\blacksquare$
\medskip

Using the equivalent representation in  Proposition~\ref{prop:Pm}, we show the following convex hull result.
\begin{Prop} \label{prop:MPSK_cvx}
    Given any integer $m \geq 3$, the set $\setP_m$ in \eqref{eq:Pm} is the convex hull of $\Theta_m$.
\end{Prop}

\medskip
{\em Proof of Proposition \ref{prop:MPSK_cvx}:} \
The set $\setP_m$ is a non-empty convex compact set;
specifically, we can  readily see from \eqref{eq:Pm} that $\setP_m$ is closed and convex, while \eqref{eq:Pm2} indicates that $\setP_m$ is non-empty and bounded.
Since $\setP_m$ is non-empty convex compact, the Krein-Milman theorem asserts that $\setP_m$ is the convex hull of the set of extreme points of $\setP_m$.
In particular, if the set of extreme points of $\setP_m$ is $\Theta_m$, then we have $\conv(\Theta_m)= \setP_m$.

To identify the extreme points of $\setP_m$, 
let $x$ be any point in $\setP_m$.
Using the representation in \eqref{eq:Pm2}, we can write 
\beq \label{eq:prop:MPSK_cvx_p1}
x =  e^{\jj \frac{2\pi k}{m}} ( \theta_2 e^{-\jj \frac{\pi}{m}} + \theta_3 e^{\jj \frac{\pi}{m}} ),
\eeq 
for some $k \in \{0,\ldots,m-1 \}$ and for some $\btheta \in \Rbb_+^3$, $\theta_1 + \theta_2 + \theta_3 =1$.
Eq.~\eqref{eq:prop:MPSK_cvx_p1} indicates that $x$ is an extreme point of $\setP_m$ only if $\btheta = \be_1$, $\btheta = \be_2$ or $\btheta = \be_3$.
For the case of $\btheta = \be_1$, the corresponding point $x= 0$ is not an extreme point of $\setP_m$.
We can write $x= 0$ as a convex combination of points in $\setP_m$; specifically, from the basic fact that $\sum_{l=0}^{m-1} e^{\jj \frac{2\pi l}{m}}= 0$, we have 
\[
0 = \sum_{l=0}^{m-1} \frac{1}{m} e^{\jj \frac{2\pi l}{m} + \jj \frac{\pi}{m}}.
\]
For the other two cases, we have $x = e^{\jj \frac{2\pi k}{m} \pm \jj \frac{\pi}{m}}$.
Suppose that $e^{\jj \frac{2\pi k}{m} \pm \jj \frac{\pi}{m}}$ is not an extreme point of $\setP_m$.
That means that there exist $x',x'' \in \setP_m$, $x' \neq x''$, and $\vartheta \in (0,1)$ such that $e^{\jj \frac{2\pi k}{m} \pm \jj \frac{\pi}{m}} = \vartheta x' + (1 - \vartheta) x''$.
From \eqref{eq:prop:MPSK_cvx_p1}, we can verify that $|x| \leq 1$ must be true for any $x \in \setP_m$.
By the Jensen inequality for strictly convex functions, we have
\[
1 = | e^{\jj \frac{2\pi k}{m} \pm \jj \frac{\pi}{m}} | < \vartheta |x'| + (1-\vartheta)|x''| \leq 1,
\]
which is a contradiction.
We hence conclude that $\Theta_m$ is the set of extreme points of $\setP_m$, and consequently we have the desired result $\conv(\Theta_m)= \setP_m$.
\hfill $\blacksquare$
\medskip

\ifconfver
\subsection{MPSK Set, Part 3: Projection}
\else
\subsubsection{MPSK Set, Part 3: Projection}
\fi

To derive the projection onto $\setP_m$, consider the following lemma.
\begin{Lemma} \label{lem:proj_cvxh}
    Given a vector $\bz \in \Rbb^n$ and a non-empty set $\setA \subseteq \Rbb^n$, consider the following problem
    \beq \label{eq:lem:proj_cvxh_1}
    \min_{\bx \in \conv(\setA)} \| \bz - \bx \|_2^2.
    \eeq 
    A point $\bx \in \conv(\setA)$ is an optimal solution to problem \eqref{eq:lem:proj_cvxh_1} if and only if 
    \beq \label{eq:lem:proj_cvxh_2}
    ( \bx - \bz )^\top (\ba - \bx ) \geq 0,
    \quad \forall \ba \in \setA.
      \nonumber
    \eeq
\end{Lemma}

\medskip
{\em Proof of Lemma \ref{lem:proj_cvxh}:} \
Problem \eqref{eq:lem:proj_cvxh_1} is convex.
As a basic result in convex optimization, a point $\bx \in \conv(\setA)$ is an optimal solution to problem \eqref{eq:lem:proj_cvxh_1} if and only if 
\beq \label{eq:lem:proj_cvxh_p1}
\nabla f(\bx)^\top (\bx' - \bx) = 2(\bx - \bz)^\top (\bx' - \bx) \geq 0, \quad \forall \bx' \in \conv(\setA).
\eeq 
Let $\psi(\bx')= (\bx - \bz)^\top (\bx' - \bx)$ for convenience.
Suppose that \eqref{eq:lem:proj_cvxh_p1} holds.
Then we have
\beq \label{eq:lem:proj_cvxh_p2}
\psi(\ba) \geq 0, \quad \forall \ba \in \setA.
\eeq 
Conversely, suppose that \eqref{eq:lem:proj_cvxh_p2} holds.
Let $\bx'$ be any point in $\conv(\setA)$, and represent it by $\bx' = \sum_{i=1}^k \theta_i \ba_i$ for some $\ba_1,\ldots,\ba_k \in \setA$, $\btheta \in \Rbb_+^k$, $\sum_{i=1}^k \theta_i = 1$, and $k$.
We have $\psi(\bx') = \sum_{i=1}^k \theta_i \psi(\ba_i) \geq 0$.
It follows that \eqref{eq:lem:proj_cvxh_p1} holds.
As a result, \eqref{eq:lem:proj_cvxh_p2} is equivalent to the optimality condition in \eqref{eq:lem:proj_cvxh_p1}.
The proof is complete.
\hfill $\blacksquare$
\medskip

The projection result is as follows.
\begin{Prop} \label{prop:proj_MPSK_cvxh}
    The projection of a given $z \in \Cbb$ onto $\setP_m$ is 
    \beq
        \Pi_{\setP_m}(z) = e^{\jj \frac{2\pi k}{m}} \left[ [\Re(y)]_0^{\cos(\pi/m)} + \jj [\Im(y) ]_{-\sin(\pi/m)}^{\sin(\pi/m)}  \right],
    \eeq 
    where $k = \left\lfloor (\angle z + \pi/m)/(2\pi/m) \right\rfloor$, $y = z e^{-\jj \frac{2\pi k}{m}}$.
\end{Prop}

\medskip
{\em Proof of Proposition \ref{prop:proj_MPSK_cvxh}:} \
To facilitate our description, let us write down the projection problem 
\beq \label{eq:prop:proj_MPSK_cvxh}
    \min_{x \in \setP_m} | z - x |^2.
    \eeq 
By Lemma \ref{lem:proj_cvxh}, a point $x \in \setP_m$ is the optimal solution to problem \eqref{eq:prop:proj_MPSK_cvxh} if and only if
\beq \label{eq:prop:proj_MPSK_cvxh_p1}
\Re( (x-z)^*(v-x) ) \geq 0, \quad \forall v \in \Theta_m.
\eeq 
We derive the optimal solution by \eqref{eq:prop:proj_MPSK_cvxh_p1}.
To do so, we employ the following representation of $z$:
\beq  \label{eq:prop:proj_MPSK_cvxh_p2}
z = e^{\jj \frac{2\pi k}{m}} y,
\eeq 
where $k \in \{0,\ldots,m-1\}$ and $y \in \Cbb$ has $\angle y \in [-\pi/m,\pi/m]$.
Also, we use Lemma~\ref{lem:mpsk2} to characterize $y$ as
\beq  \label{eq:prop:proj_MPSK_cvxh_p25}
y  = \alpha  \left[ \cos\left( \tfrac{\pi}{m} \right) + \jj \beta \sin\left( \tfrac{\pi}{m} \right) \right],
\eeq 
for some $\alpha \geq 0$, $\beta \in [-1,1]$.
We divide the proof into several cases.

{\em Case (a): $\alpha \leq 1$.}
We have $z \in \setP_m$; see Proposition \ref{prop:Pm} and Lemma~\ref{lem:mpsk1}.
Choosing $x = z$ leads to the satisfaction of the optimality condition \eqref{eq:prop:proj_MPSK_cvxh_p1}.

{\em Case (b): $\alpha > 1$, $\alpha |\beta| \leq 1$.}
Choose
\beq \label{eq:prop:proj_MPSK_cvxh_p3}
x = e^{\jj \frac{2\pi k}{m}} \left[ \cos\left( \tfrac{\pi}{m} \right) + \jj \alpha \beta \sin\left( \tfrac{\pi}{m} \right) \right].
\eeq 
This $x$ lies in $\setP_m$; see Proposition \ref{prop:Pm} and Lemma~\ref{lem:mpsk1}.
We have $x - z  = -r e^{\jj \frac{2\pi k}{m}}$, where $r = (\alpha - 1) \cos(\pi/m) > 0$. 
Applying the above equation to the consitutent terms of \eqref{eq:prop:proj_MPSK_cvxh_p1} gives
\begin{align*}
    \Re( (x-z)^* x ) & = -r \cos\left( \tfrac{\pi}{m} \right);  \\
    \intertext{and, for any $v = e^{\jj \frac{2\pi l}{m} + \jj \frac{\pi}{m}}$, $l \in \{0,\ldots,m-1\}$,}
    \Re( (x-z)^* v ) & = -r \cos\left(  \tfrac{2\pi (l-k)}{m} + \tfrac{\pi}{m} \right) \geq - r \cos\left( \tfrac{\pi}{m} \right),
\end{align*}
where the inequality is due to Lemma \ref{lem:mpsk3}.
We see that the optimality condition \eqref{eq:prop:proj_MPSK_cvxh_p1} are satisfied by the $x$ in \eqref{eq:prop:proj_MPSK_cvxh_p3}.

{\em Case (c): $\alpha > 1$, $\beta > 0$, $\alpha \beta > 1$.}
Choose
\beq \label{eq:prop:proj_MPSK_cvxh_p4}
x = e^{\jj \frac{2\pi k}{m}} \left[ \cos\left( \tfrac{\pi}{m} \right) + \jj  \sin\left( \tfrac{\pi}{m} \right) \right] \in \setP_m.
\eeq 
We have $x - z =  -h e^{\jj \frac{2\pi k}{m}}$, where 
\[
h =  (\alpha - 1) \cos\left( \tfrac{\pi}{m} \right) + \jj  (\alpha\beta - 1)  \sin\left( \tfrac{\pi}{m} \right). 
\]
We represent $h$ by the polar form $h = r e^{\jj \phi}$.
Particularly, $\phi$ is given by
\[
\phi = \tan^{-1} \left(  \frac{\alpha \beta -1}{\alpha - 1} \tan\left( \tfrac{\pi}{m}  \right) \right). 
\]
Since $(\alpha \beta -1)/(\alpha - 1) \in (0,1]$, it can be verified that $\phi \in (0, \pi/m]$; the argument is identical to a key argument in the proof of Lemma~\ref{lem:mpsk2}.
Using the above representation of $x - z$, we get
\ifconfver
\begin{align*}
    \Re( (x-z)^* x ) & = -r \cos\left( \tfrac{\pi}{m} - \phi \right);  \\
    \intertext{and, for any $v = e^{\jj \frac{2\pi l}{m} + \jj \frac{\pi}{m}}$, $l \in \{0,\ldots,m-1\}$,}
    \Re( (x-z)^* v ) & = -r \cos\left(  \tfrac{2\pi (l-k)}{m} + \tfrac{\pi}{m} - \phi \right)\\ 
    &\geq - r \cos\left( \tfrac{\pi}{m} - \phi \right),
\end{align*}
\else
\begin{align*}
    \Re( (x-z)^* x ) & = -r \cos\left( \tfrac{\pi}{m} - \phi \right);  \\
    \intertext{and, for any $v = e^{\jj \frac{2\pi l}{m} + \jj \frac{\pi}{m}}$, $l \in \{0,\ldots,m-1\}$,}
    \Re( (x-z)^* v ) & = -r \cos\left(  \tfrac{2\pi (l-k)}{m} + \tfrac{\pi}{m} - \phi \right) 
    \geq - r \cos\left( \tfrac{\pi}{m} - \phi \right),
\end{align*}
\fi
where the inequality is due to Lemma \ref{lem:mpsk3}.
The optimality condition \eqref{eq:prop:proj_MPSK_cvxh_p1} is seen to be satisfied by the $x$ in \eqref{eq:prop:proj_MPSK_cvxh_p4}.

{\em Case (d): $\alpha > 1$, $\beta < 0$, $\alpha \beta < -1$.} It can be shown that 
\beq 
x = e^{\jj \frac{2\pi k}{m}} \left[ \cos\left( \tfrac{\pi}{m} \right) - \jj  \sin\left( \tfrac{\pi}{m} \right) \right] \in \setP_m
\eeq 
satisfies the optimality condition \eqref{eq:prop:proj_MPSK_cvxh_p1}.
The proof is almost the same as that of Case (c), and we shall omit the details.

In the above we have shown the solution to problem \eqref{eq:prop:proj_MPSK_cvxh} for different cases.
Assembling the results together, we can write down the solution as a single formula
\beq \label{eq:prop:proj_MPSK_cvxh_p5}
x = e^{\jj \frac{2\pi k}{m}} \left[ [\alpha]_0^1 \cos\left( \frac{\pi}{m} \right) - \jj [\alpha\beta]_{-1}^1 \sin\left( \frac{\pi}{m} \right) \right].
\eeq 
What remains to show is how the parameters $k$, $\alpha$ and $\beta$ are obtained from $z$.
From \eqref{eq:prop:proj_MPSK_cvxh_p2}, it can be shown that 
\[
k = \left\lfloor \frac{\angle z + \frac{\pi}{m}}{\frac{2\pi}{m}} \right\rfloor.
\]
Also, from \eqref{eq:prop:proj_MPSK_cvxh_p25}, we have
\[
\Re(z e^{-\jj \frac{2\pi k}{m}}) = \alpha \cos\left( \tfrac{\pi}{m} \right),
\quad 
\Im( z e^{-\jj \frac{2\pi k}{m}}) = \alpha \beta \sin\left( \tfrac{\pi}{m} \right).
\]
Applying the above results to \eqref{eq:prop:proj_MPSK_cvxh_p5} leads to the desired result.

\hfill $\blacksquare$
\medskip

\ifconfver

\else
\bibliographystyle{ieeetr}
\bibliography{refs,refs_JB_part1}
\fi

\end{document}

%% file: sym_marco.tex
\newcommand\bw{\ensuremath{{\bm w}}}
\newcommand\bW{\ensuremath{{\bm W}}}

\newcommand\bx{\ensuremath{{\bm x}}}
\newcommand\by{\ensuremath{{\bm y}}}

\newcommand\bH{\ensuremath{{\bm H}}}

\newcommand\be{\ensuremath{{\bm e}}}
\newcommand\bz{\ensuremath{{\bm z}}}

\newcommand\bX{\ensuremath{{\bm X}}}
\newcommand\bZ{\ensuremath{{\bm Z}}}

\newcommand\ba{\ensuremath{{\bm a}}}

\newcommand\bA{\ensuremath{{\bm A}}}

\newcommand\bb{\ensuremath{{\bm b}}}

\newcommand\bB{\ensuremath{{\bm B}}}

\newcommand\bd{\ensuremath{{\bm d}}}
\newcommand\bD{\ensuremath{{\bm D}}}

\newcommand\bv{\ensuremath{{\bm v}}}
\newcommand\bSig{\ensuremath{{\bm \Sigma}}}
\newcommand\bsig{\ensuremath{{\bm \sigma}}}

\newcommand\btheta{\ensuremath{{\bm \theta}}}

\newcommand\bkappa{\ensuremath{{\bm \kappa}}}
\newcommand\bka{\ensuremath{{\bm \kappa}}}
\newcommand\ka{\ensuremath{{\kappa}}}

\newcommand\bY{\ensuremath{{\bm Y}}}
\newcommand\bV{\ensuremath{{\bm V}}}
\newcommand\bU{\ensuremath{{\bm U}}}

\newcommand{\Rbb}{\mathbb{R}}
\newcommand{\Cbb}{\mathbb{C}}

\newcommand{\Zbb}{\mathbb{Z}}

\newcommand{\setA}{\mathcal{A}}
\newcommand{\setB}{\mathcal{B}}
\newcommand{\setD}{\mathcal{D}}

\newcommand{\setX}{\mathcal{X}}

\newcommand{\setU}{\mathcal{U}}
\newcommand{\setV}{\mathcal{V}}
\newcommand{\setW}{\mathcal{W}}
\newcommand{\setS}{\mathcal{S}}
\newcommand{\setT}{\mathcal{T}}
\newcommand{\setC}{\mathcal{C}}
\newcommand{\setN}{\mathcal{N}}

\newcommand{\setP}{\mathcal{P}}

\newcommand{\jj}{\mathfrak{j}}

\newcommand{\Diag}{\mathrm{Diag}}

\newcommand{\eps}{\varepsilon}
\newcommand{\bzero}{{\bm 0}}
\newcommand{\bone}{{\bm 1}}
\newcommand{\bI}{{\bm I}}

\newcommand\bigO{\ensuremath{{\mathcal{O}}}}
\newcommand\half{\ensuremath{{\frac{1}{2}}}}

\newcommand\conv{\ensuremath{{\rm conv}}}

\newcommand\tr{\ensuremath{{\rm tr}}}

\newcommand\dist{\ensuremath{{\rm dist}}}
\newcommand\fro{\ensuremath{{\rm F}}}